\documentclass[twocolumn,showpacs,aps,prb,superscriptaddress]{revtex4-2}
\usepackage{xcolor}
\usepackage[colorlinks=true, linkcolor=blue, urlcolor=magenta, citecolor=red]{hyperref}
\usepackage{amssymb}
\usepackage{graphicx}
\usepackage{amsmath}
\usepackage{color}
\usepackage[caption=false]{subfig}
\usepackage{slashed}
\usepackage{float}
\renewcommand{\b}[1]{\boldsymbol{#1}}

\newcommand{\Tr}{\text{Tr}}
\newcommand{\tr}{\text{tr}}
\newcommand{\N}{\mathcal{N}}

\renewcommand{\k}{\b{k}}

\renewcommand{\c}[1]{\mathcal{#1}}

\newcommand{\nn}{\nonumber}

\def\beq#1\eeq{\begin{equation}\begin{aligned}#1\end{aligned}\end{equation}}

\newcommand{\lb}{\nonumber\\}

\graphicspath{{./}{./figs/}}

\definecolor{JM}{rgb}{0,0.5,0}

\begin{document}

\title{
Field-driven transition from quantum spin liquid to magnetic order in triangular-lattice antiferromagnets
}
\date{\today}

\author{Santanu Dey}
\affiliation{Department of Physics, University of Alberta, Edmonton, Alberta, T6G 2E1, Canada}
\author{Joseph Maciejko}
\affiliation{Department of Physics, University of Alberta, Edmonton, Alberta, T6G 2E1, Canada}
\affiliation{Theoretical Physics Institute (TPI), University of Alberta, Edmonton, Alberta, T6G 2E1, Canada}
\author{Matthias Vojta}
\affiliation{Institut f\"ur Theoretische Physik and W\"urzburg-Dresden Cluster of Excellence ct.qmat, Technische Universit\"at Dresden, 01062 Dresden, Germany}


\begin{abstract}
Recently, several triangular-lattice magnets with delafossite structure have been found to display spin-liquid behavior down to the lowest temperatures. Remarkably, applying a magnetic field destroys the spin liquid which then gives way to symmetry-breaking states, identified as semiclassical coplanar states including a magnetization plateau at 1/3 total magnetization.
Here we provide a theoretical approach rationalizing this dichotomy, utilizing a Schwinger-boson theory that captures both ordered and disordered magnetic phases. We show that a zero-field spin liquid, driven by strong frustration, is naturally destabilized in a magnetic field via spinon condensation. Symmetry-breaking order akin to the standard triangular-lattice Heisenberg model then arises via an order-by-disorder mechanism.
We discuss implications for pertinent experiments.
\end{abstract}

\maketitle

\section{Introduction}
\label{intro}

Frustrated interactions in local-moment magnets tend to suppress magnetic order and can lead to low-temperature states
defying a description in terms of symmetry-breaking order parameters and their fluctuations~\cite{balents2010,savary16,zhou2017}. 
There is significant interest in studying magnetic compounds which,
by means of a control parameter such as
chemical substitution or applied magnetic field,
display transitions between magnetically ordered phases and paramagnetic quantum spin-liquid (QSL) phases. 
A theoretical account of such transitions necessarily lies beyond the Landau symmetry-breaking paradigm and requires concepts such as fractionalization and long-range entanglement~\cite{WenBook}. 

Recent experimental studies of 
rare-earth delafossites~\cite{baenitz18,sarkar2019,ding19,bordelon19,ranjith19,bordelon2020,ranjith19b,xing19}, compounds of the form $A^{1+}R^{3+}X_2$ with $A$ a nonmagnetic ion, $R$ a rare-earth ion, and $X$ a chalcogen, suggest that these compounds are an ideal platform for investigating the competition between QSL and magnetically ordered ground states, and how it unfolds in the presence of an external magnetic field. These layered compounds feature $j_\text{eff}=1/2$ moments on a structurally perfect triangular lattice and are believed to be exceptionally clean. While some of them, such as KCeS$_2$~\cite{bastien20}, have been found to display magnetic order at low temperatures, there is an entire family of QSL candidates, encompassing NaYbS$_2$~\cite{baenitz18,sarkar2019}, NaYbO$_2$~\cite{ding19,bordelon19,ranjith19,bordelon2020}, NaYbSe$_2$~\cite{ranjith19b}, and CsYbSe$_2$~\cite{xing19,xie21}, where no zero-field order has been detected; KYbSe$_2$ displays weak order at zero field but is argued to be near a QSL quantum critical point~\cite{scheie21,scheie22}. Remarkably, upon application of a magnetic field, these compounds display a sequence of ordered phases identified as coplanar three-sublattice states including a magnetization plateau at $1/3$ of the saturation magnetization~\cite{bordelon19,ranjith19,bordelon2020,ranjith19b,xing19,xing21}.

This sequence of field-induced ordered phases is reminiscent of the zero-temperature phase diagram of the nearest-neighbor triangular-lattice Heisenberg antiferromagnet (TLHAF), obtained, e.g., using the semiclassical $1/S$ expansion by Chubukov and Golosov~\cite{chubukov91}. At zero field, the TLHAF orders in a non-collinear $120^\circ$ three-sublattice magnetic spiral. With an applied magnetic field, the spiral is replaced in the classical $S\rightarrow\infty$ limit by a degenerate ground-state manifold involving coplanar and non-coplanar states. Beyond the classical limit, a quantum order-by-disorder mechanism~\cite{henley89} lifts this accidental degeneracy in favor of coplanar three-sublattice states, including the so-called Y state at low fields, a 2:1 canted (V) state at high fields, and an incompressible 1/3 magnetization plateau (up-up-down state) at intermediate fields. 
This complex interplay between geometric frustration and applied magnetic field in the TLHAF has invited significant experimental activity over the years~\cite{fortune09,ono11,shirata12,kamiya18}, 
culminating in the recent surge of interest in delafossite compounds. The new ingredient in those latest experiments is the absence of $120^\circ$ order at low fields in favor of a QSL state, which calls for theoretical treatments beyond the semiclassical paradigm that can adequately capture the observed field-induced competition between fractionalization and conventional symmetry breaking.

Motivated by these questions, we explore theoretically the competition between QSL physics and field-induced magnetic order on the triangular lattice using Schwinger-boson methods~\cite{arovas88,read91,sachdev91,sachdev92,auerbach94,yoshioka91}. We consider the TLHAF model in an applied magnetic field $\b{h}$,
\begin{align}
  \mathcal{H} =
  \sum_{(ij)} J_{ij} \b{S}_i\cdot\b{S}_j 
  - S\b{h} \cdot \sum_i \b{S}_i,
  \label{heisen_zeeman}
\end{align}
where $\b{S}_i$ is the spin operator on site $i$ and $J_{ij}>0$ is the antiferromagnetic exchange coupling between sites $i$ and $j$. For describing the magnetism of the delafossites, both spin-isotropic models \cite{scheie21,scheie22} with first-neighbor ($J_1$) and second-neighbor ($J_2$) couplings, with $J_2/J_1$ acting as a frustration parameter, as well as various spin-anisotropic models \cite{cherny18,cherny19,pocs21,schmidt21} have been considered. For simplicity we shall ignore possible spin anisotropies. Instead of studying the $J_1$-$J_2$ model directly, we will consider nearest-neighbor ($J$) exchange only but use a parameter $\kappa$ controlling the spin size $S$ as a proxy for the (inverse) strength of quantum fluctuations. As detailed in Sec.~\ref{largeN}, we enlarge the spin symmetry from SU(2) to Sp(N) to develop a controlled theory in the large-$N$ limit~\cite{read91,sachdev91,sachdev92}. We then extrapolate our results to the physical $N=1$ limit, corresponding to SU(2) spins and for which $\kappa=2S$ with $S$ the spin quantum number.

\begin{figure}
\includegraphics[width=1\columnwidth]{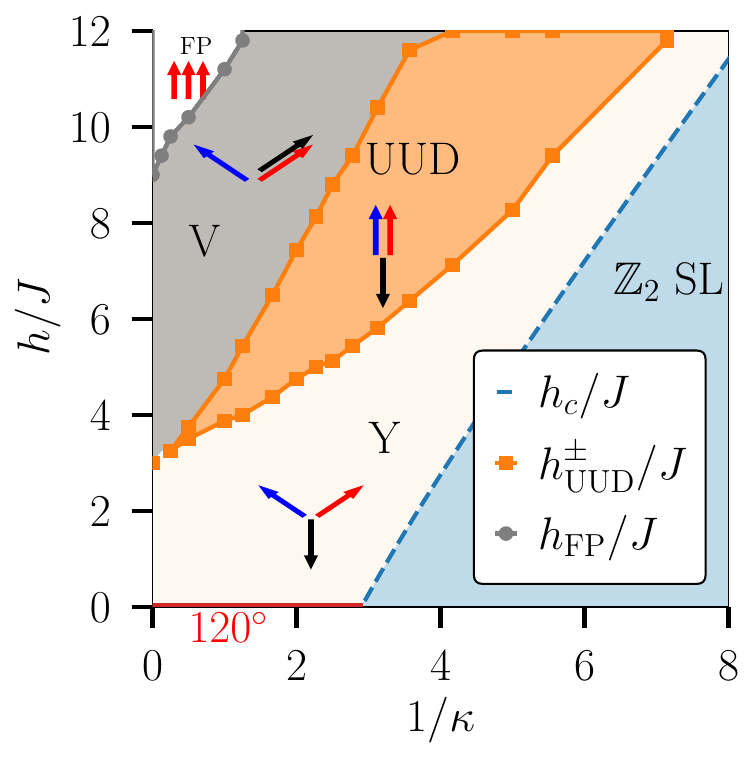}
\caption{
Phase diagram of the triangular-lattice Heisenberg antiferromagnet obtained from 
  the Sp(N) Schwinger-boson theory, 
  as a function of the magnetic field $h$ 
    and the inverse spin-size $1/\kappa$, with 
  $\kappa=\infty$ describing the classical limit.
  A $\mathbb{Z}_2$ quantum spin liquid (QSL) 
  transitions to non-collinear 120$^\circ$ order at zero field (red line) and a sequence of canted antiferromagnetic phases at a critical field $h_c(\kappa)$ (dashed blue line). The ordered phases include the semiclassical coplanar Y and V states and an up-up-down (UUD) 1/3 magnetization plateau phase. Beyond a field $h_{\rm FP}$ a fully-polarized state (FP) is obtained. Since the coplanar ordered phases are favored over non-coplanar ones only by nominally $1/N$ fluctuation corrections, the phase boundaries of these phases have been obtained including such corrections. 
}
\label{fluctPd}
\end{figure}

Our main results are depicted in the zero-temperature phase diagram, Fig.~\ref{fluctPd}, as a function of the strength $1/\kappa$ of quantum fluctuations and the magnitude $h$ of the applied field, and they can be summarized as follows. At zero field, the ground state is a gapped $\mathbb{Z}_2$ QSL for large $1/\kappa$ and a magnetic state with 120$^\circ$ non-collinear order for small $1/\kappa$; the latter is obtained from condensation of the gapped bosonic spinons in the QSL at a critical value of the inverse spin-size~\cite{sachdev92}. For small but finite $1/\kappa$, we find that turning on an external field $h$ reproduces the same sequence of ordered states as the semiclassical treatment of the TLHAF~\cite{chubukov91}. Crucially, we find the strict large-$N$ limit leads to a near degeneracy between coplanar and non-coplanar states. We then compute $1/N$ corrections to the ground-state energy, adapting the recently proposed formalism of References~\onlinecite{ghioldi18,zhang19,zhang22} to the case of finite magnetic fields, and we find that such corrections favor the coplanar states. For $1/\kappa$ larger than its zero-field critical value, we find that the $\mathbb{Z}_2$ QSL becomes unstable at a critical magnetic field value $h_c(\kappa)$ beyond which magnetic order sets in via spinon condensation. The transition out of the QSL is found to be continuous and results in the Y state. At some higher magnetic field $h_1^\text{UUD}$, we find a first-order transition to a collinear 1/3 magnetization plateau state. This state persists until $h_2^\text{UUD}>h_1^\text{UUD}$, beyond which we find another first-order transition to a canted V state. This state undergoes a final continuous transition into the fully polarized state beyond a saturation field $h_\text{FP}$.

The rest of the paper is structured as follows. In Sec.~\ref{largeN}, we introduce the Sp(N) generalization of the TLHAF Hamiltonian and its representation using Schwinger bosons. We solve the mean-field equations of the large-$N$ limit and obtain a phase diagram with QSL and ordered phases corresponding to uncondensed or condensed Schwinger bosons, respectively. Among ordered phases, we find a near degeneracy between coplanar and non-coplanar states. This ambiguity is resolved by considering $1/N$ corrections in Sec.~\ref{fluctuation}; they favor coplanar states and produce the sequence of ordered states expected for the TLHAF. $1/N$ corrections are also necessary to obtain a dynamical spin structure factor $S(\b{p},\omega)$ with the correct magnon physics in the ordered phases~\cite{ghioldi18,zhang19,zhang22}. Finally, we discuss our results in the context of recent experiments on delafossite compounds, and we comment on future extensions of our work.

\section{Large-$N$ Schwinger Boson Theory}
\label{largeN}

\subsection{Model and mean-field equations}

We begin with the TLHAF Hamiltonian (\ref{heisen_zeeman}), which we study using a representation of spin operators in terms of Schwinger bosons $b_\alpha,b^\dag_\alpha$ with $\alpha=\uparrow,\downarrow$~\cite{auerbach94}; these are interpreted physically as bosonic spinon degrees of freedom. The SU(2) spin operator on site $i$ is expressed as $\b{S}_i = \frac{1}{2} b^\dagger_{i\alpha}\b{\sigma}_{\alpha\beta}b_{i\beta}$, with the physical spin-$S$ Hibert space imposed by the constraint $b_{i\alpha}^\dagger b_{i\alpha}=2S$. To obtain a controlled theory using large-$N$ methods, one must enlarge the SU(2) symmetry group. For
bipartite lattices,
$SU(N)$ generalizations have been put forward for which mean-field theory becomes exact in the limit 
$N\to\infty$~\cite{arovas88,auerbach94}. Here we employ a symplectic Sp(N) generalization suitable for 
non-bipartite lattices~\cite{read91,sachdev91,sachdev92}, where the mean-field theory involves a decoupling in the singlet particle-particle channel and hence a BCS-type ground state. 
Using Schwinger bosons $b_{ia\alpha},b_{ia\alpha}^\dag$ with flavor index $a=1,\ldots,N$ and the local constraint
\begin{align}\label{constraint}
\sum_{a\alpha} b_{ia\alpha}^\dagger b_{ia\alpha} = \kappa N,
\end{align}
the Sp(N) generalization of Eq.~\eqref{heisen_zeeman} can be written as \cite{sachdev92}
\begin{equation}
  \begin{aligned}
\mathcal{H} =
  &-\sum_{(ij)}\frac{J_{ij}}{2N}
  \left[
    \left(\mathcal E^{ab}_{\alpha\beta}b^\dagger_{ia\alpha} b^\dagger_{jb\beta}\right)
    \left(\mathcal E^{cd}_{\gamma\delta}b_{ic\gamma} b_{jd\delta}\right)
    -\frac{N^2 \kappa^2}{2}
    \right]\\
  &    -\frac{h\kappa}{4}\sum_i b^\dagger_{ia\alpha}\sigma^z_{\alpha\beta}b_{ia\beta},
  \end{aligned}
  \label{largeN_ham}
\end{equation}
with the symplectic tensor
$\mathcal E^{ab}_{\alpha\beta} = \delta^{ab} \epsilon_{\alpha\beta}$ appearing as
a large-$N$ extension of the Levi-Civita symbol, and we have chosen the field direction along the $z$ axis without loss of generality. We note that the external field couples equally to all $N$ flavor pairs. On account of the group isomorphism $Sp(1)\cong SU(2)$, Eq.~(\ref{largeN_ham}) is equivalent to the SU(2) TLHAF \eqref{heisen_zeeman} when $N=1$ and $\kappa=2S$.

Expressing the partition function of the Sp(N) theory (\ref{largeN_ham}) as an imaginary-time functional integral over Schwinger boson fields, we further decompose the four-boson interaction term via complex Hubbard-Stratonovich fields $Q_{ij}(\tau),Q_{ij}^*(\tau)$ to obtain the action:
\begin{align}
    \mathcal S[b,b^*,Q,Q^*,\lambda] &=\int\!d\tau\Big( 
    \sum_{i}b_{ia\alpha}^*\partial_\tau b_{ia\alpha} +\colon\mathcal H_{\rm MF}[b,b^*]\colon\Big),\label{SBact}
\end{align}
with the effective Hamiltonian \footnote{Alternative large-$N$ treatments involve both particle-particle ($b_ib_j,b_i^\dag b_j^\dag$) and particle-hole ($b_i^\dag b_j$) bilinears~\cite{wang06,flint09}, but we will not consider those here.
}
\begin{align}
    &\mathcal{H}_{\rm MF}[b,b^\dag]\nn\\
    &\hspace{10mm}= \frac{1}{2} \sum_{(ij)}J_{ij}
  \Big[
    N |Q_{ij}|^2 - \big( Q_{ij} \epsilon_{\sigma\sigma'} b^\dagger_{ia\alpha} b^\dagger_{ja\beta} \!+\! \mathrm{h.c.} \big)
  \Big] \nn\\
  &\hspace{10mm}- \frac{1}{2}\left(\frac{h\kappa}{2}+B_z\right)\sum_{i} b_{ia\alpha}^\dagger \sigma^z_{\alpha\beta} b_{ia\beta}\nn\\
 &\hspace{10mm} + i\sum_i \lambda_i \big( b_{ia\alpha}^\dagger b_{ia\alpha} - \kappa N \big) ,
  \label{Heff}
\end{align}
where the constant piece $\propto\kappa^2$ has been dropped.
A Lagrange multiplier $\lambda_i(\tau)$ has been utilized to impose the local constraint \eqref{constraint} in the partition sum, and we have introduced a probe field $B_z$ (distinct from the applied field $h$) for eventual computations of the uniform magnetization. In the functional-integral representation, in which a systematic $1/N$ expansion can be formulated, the bosonic operators $b_{ia\alpha},b_{ia\alpha}^\dag$ are replaced by complex fields $b_{ia\alpha}(\tau),b_{ia\alpha}^*(\tau)$ in the normal-ordered Hamiltonian $\colon\mathcal H_{\rm MF}\colon$. 

The formal large-$N$ expansion is obtained after integrating out the quadratic bosonic fields arranged in a Nambu basis 
$\Phi_{a}(\tau)=\begin{pmatrix}\dots, b_{ia\uparrow}(\tau), \dots &&\dots, b_{ia\downarrow}(\tau)^*,\dots
\end{pmatrix}^T$ so that the partition function
assumes a simple form,
\beq
\mathcal Z = &\int \c{D}Q\c{D}Q^*\c{D}\lambda\,
e^{-N\mathcal W(\hat Q,\hat Q^\dagger,\hat\lambda)},
\label{actPart}
\eeq
where $\c{W}(\hat Q,\hat Q^\dagger,\hat\lambda)$ is the action functional obtained by integrating out the bosonic spinors from  
the Schwinger-boson action $\c{S}$ [Eq.~\eqref{SBact}]: 
\beq
\mathcal W(\hat Q,\hat Q^\dagger,\hat\lambda)=&
  \frac{1}{2}\int d\tau\sum_{(ij)}J_{ij}|Q_{ij}(\tau)|^2\\
&-i\int d\tau\sum_i \lambda_i(\tau)(\kappa+1)
\\
&+\frac{1}{N} \Tr\ln\hat G^{-1}(\hat Q,\hat Q^*,\hat\lambda),
\eeq
with a trace ($\Tr$) over the spatio-temporal, flavor, 
and spinor indices of the inverse Schwinger-boson Green's function:
\beq
&[\hat G^{-1}(\hat Q,\hat Q^*,\hat\lambda)]_{ab}\\
&\hspace{20mm}=
\delta_{ab}\Bigg[\left(\partial_\tau-\frac{1}{2}\left(\frac{h\kappa}{2}+B_z\right)\right)\hat I
\otimes\hat \sigma^z\\
&\hspace{20mm}+\begin{pmatrix}
  i\hat\lambda &&-\frac{1}{2}\hat J\circ\hat Q\\
  -\frac{1}{2}\hat J\circ \hat Q^\dagger&&i\hat\lambda
\end{pmatrix}\Bigg],
\eeq
where $\hat I$ is the $\N\times\N$ identity matrix, with $\N$ the number of lattice sites, and 
$[\hat J]_{ij} = J_{ij}$, $[\hat Q]_{ij} = Q_{ij}$, and $[\hat\lambda]_{ij}=\delta_{ij}\lambda_i$
are all expressed
as $\N\times\N$ matrices 
with $\circ$ denoting
the Hadamard product between the exchange coupling matrix and the Hubbard-Stratonovich
fields.
With the control parameter $1/N$, the partition function can be treated perturbatively by expanding around
its saddle-point value.

The saddle point itself, however, is most easily discussed in the canonical Hamiltonian framework with 
Eq.~\eqref{Heff}.
The saddle-point solution in the disordered phase is described by 
propagating Schwinger bosons and the ordered phases are obtained from the 
Bose-Einstein condensates (BEC) of these fractionalized 
quasiparticles~\cite{hirsch1989,sarker1989,chandra90} (Sec.~\ref{sec:BEC}). 
In the $N\rightarrow\infty$ limit, the saddle-point approximation to the 
partition function
becomes exact. 
Thus, the large-$N$ phase diagram is obtained by solving the time-independent saddle-point equations:
\begin{equation}
  \begin{aligned}
    \frac{\delta\mathcal S}{\delta Q^*_{ij}}=0&\Leftrightarrow Q_{ij}  = \frac{1}{N}\epsilon_{\alpha\beta}
\langle b_{ia\alpha}b_{ja\beta}\rangle,\\
    \frac{\delta\mathcal S}{\delta\lambda_i}
    =0&\Leftrightarrow \kappa=\frac{1}{N}\langle b_{ia\alpha}^\dagger b_{ia\alpha}\rangle,
  \end{aligned}
  \label{eom_master}
\end{equation}
which determine the magnitude of the oriented bond-order parameter $Q_{ij}= -Q_{ji}$ and impose the local constraint on 
average, respectively. In a putative condensate phase 
$\langle b_{ia\alpha}\rangle\neq 0$, these equations are complemented by a 
third set of equations
$\delta\mathcal S/\delta b_{ia\alpha}^*=0$ that determine the wave function of the 
possible BEC $z_{ia\alpha}\equiv\langle b_{ia\alpha}\rangle$. 
To capture ordered states with a three-sublattice ($A,B,C$) structure, we look for static solutions with three bond variables $Q_{AB},Q_{BC},Q_{CA}$, local chemical potentials $\mu_A,\mu_B,\mu_C$ such that $\lambda_i\equiv-i\mu_i$, and condensate amplitudes $z_{A\alpha},z_{B\alpha},z_{C\alpha}$, which are in general distinct. Solving the saddle-point equations ensures one has found an extremum in the free-energy density. We focus primarily on $T=0$, and to determine the global phase diagram we choose the solution with the lowest energy density.

\subsection{Uncondensed spinons: $\mathbb{Z}_2$ quantum spin liquid}

We first consider the simplest type of solution, consisting of uncondensed Schwinger bosons $\langle b_{ia\alpha}\rangle=0$. In this case, the equation $\delta\mathcal S/\delta b_{ia\alpha}^*=0$ is trivially solved and the remaining equations (\ref{eom_master}) determine the values of $Q_{ij}$ and $\mu_i$. Following Ref.~\cite{sachdev92}, we consider a homogeneous ansatz $Q_{AB}=Q_{BC}=Q_{CA}\equiv Q$, $\mu_A=\mu_B=\mu_C\equiv\mu$ such that the effective Hamiltonian (\ref{Heff}) describes nearest-neighbor spinon hopping on the triangular lattice. Passing to the Fourier domain (Appendix~\ref{app:3sub}), $\mathcal{H}_{\rm MF}$ can be diagonalized by a Bogoliubov transformation, which allows us to find the spinon excitation spectrum:
\begin{align}
    \omega_\pm(\b{k})=\sqrt{\mu^2-J^2Q^2\Bigl(\sum_{j=1,2,3}\sin\b{k}\cdot\b{u}_j\Bigr)^2}\pm\frac{|h|\kappa}{4},
\end{align}
where $\b{k}$ is the wave vector and $\b{u}_j$, $j=1,2,3$ are nearest-neighbor vectors on the triangular lattice [Fig.~\ref{lattice}(a)]. Physically, this state is a QSL with gapped bosonic spinons and $\mathbb{Z}_2$ topological order. The Schwinger bosons remain uncondensed provided the spinon gap
\begin{align}\label{gap_largen}
    \Delta=\sqrt{\mu^2-(27/4)J^2Q^2}-h\kappa/4,
\end{align}
remains real and nonnegative.

\begin{figure}
  \subfloat[]{
    \includegraphics[width=0.47\columnwidth]{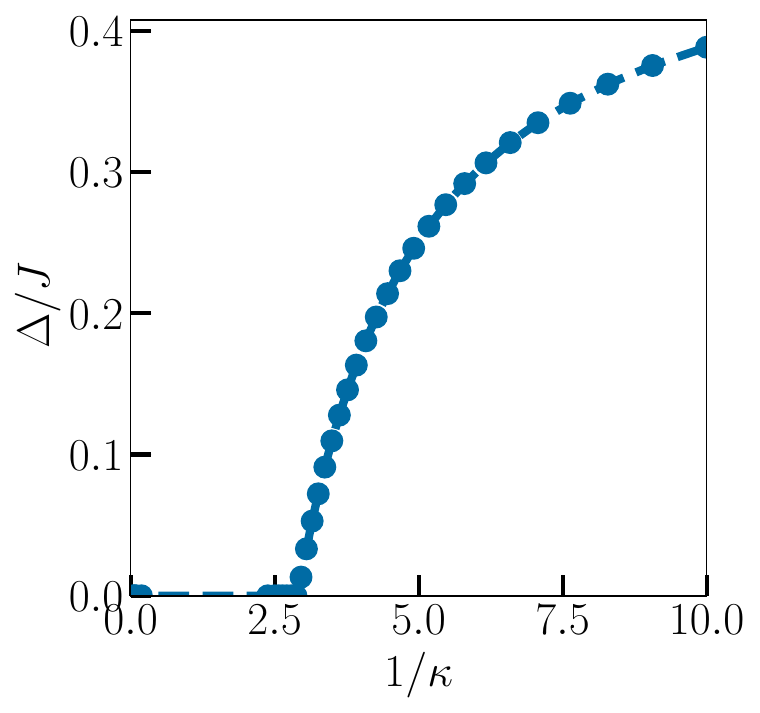}
  }
  \hfill
  \subfloat[]{
    \includegraphics[width=0.45\columnwidth]{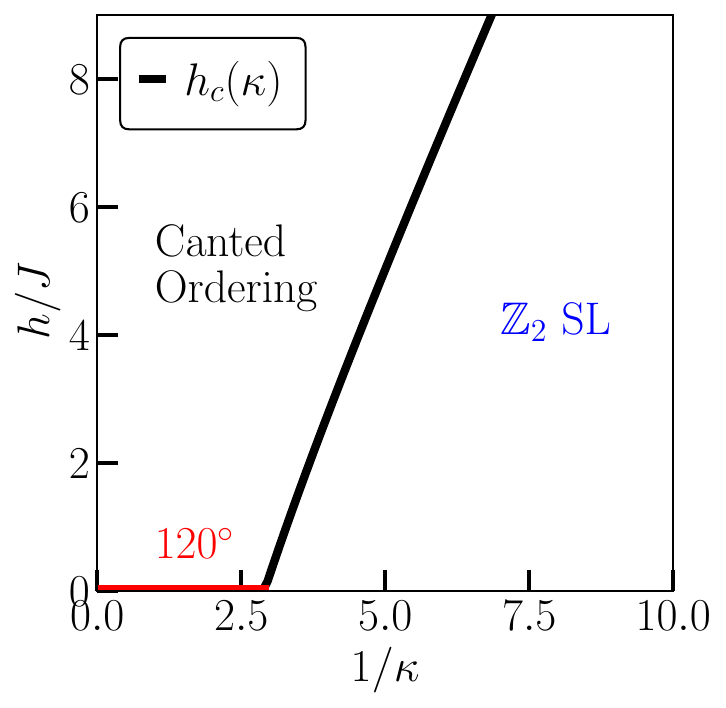}
  }
  \caption{Stability of the $\mathbb{Z}_2$ QSL in large-$N$ Schwinger-boson theory: (a) zero-field spinon gap in the QSL phase; (b) critical magnetic field 
  $h_c(\kappa)$ for spinon-gap closing towards a canted broken-symmetry phase.
  }
\label{fig:gap}
\end{figure}

At zero field, this $\mathbb{Z}_2$ QSL is the lowest-energy saddle-point solution for $\kappa<\kappa_c$ with $\kappa_c\sim 0.34$~\cite{sachdev92}. This zero-flux state yields a lower energy compared to other QSL states such as a homogeneous $\pi$-flux state~\cite{wang06}; flux corresponds here to the circulation ${\rm arg}(\prod_{(ij)\in \diamond }Q_{ij})$ of the bond order parameter around the rhombus formed by two adjacent triangles on the triangular lattice [see the shaded rhombus in Fig.~\ref{lattice}(a)]. This outcome is also consistent with the flux-expulsion principle
for symmetric and uniform bosonic spin liquids ($\lambda_i=\lambda$) in nearest-neighbor Sp(N) models~\cite{tchernyshyov06}. At $\kappa=
\kappa_c$, the spinon gap closes [Fig.~\ref{fig:gap}(a)] at wave vectors that indicate translation symmetry breaking and a tripling of the unit cell. For $\kappa>\kappa_c$, one obtains a spinon BEC corresponding to 120$^\circ$ non-collinear order~\cite{sachdev92,yoshioka91}.

\subsection{Spinon BEC: ordered phases}\label{sec:BEC}

\begin{figure}
  \subfloat[]{
    \includegraphics[width=0.95\columnwidth]{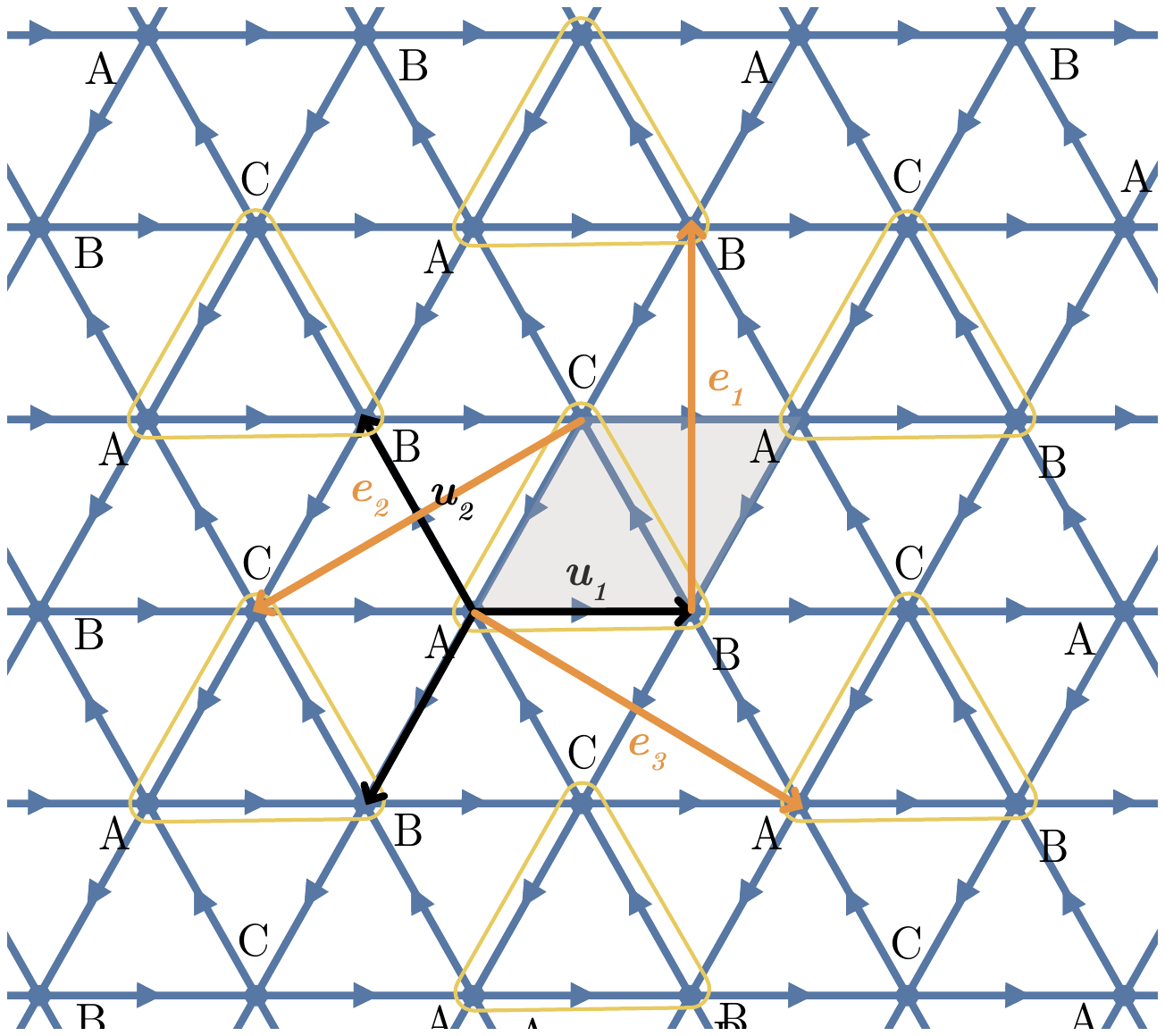}
  }
  \hspace{0pt}
  \subfloat[]{
    \includegraphics[width=0.95\columnwidth]{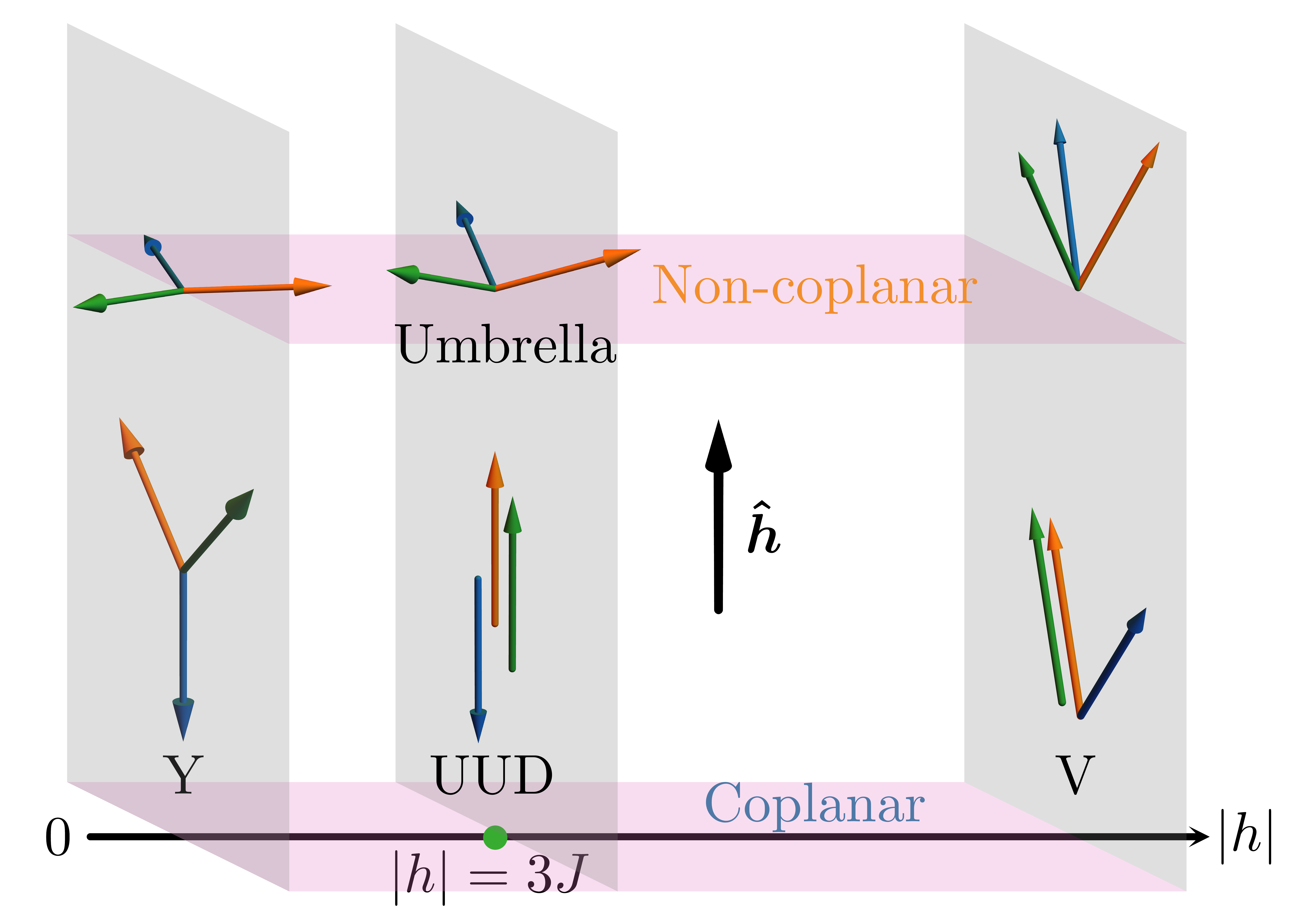}
  }
  \caption{(a) Triangular lattice with nearest-neighbor vectors $\b{u}_{1,2,3}$ (black arrows), and oriented links (blue arrows) depicting the bond order parameters $Q_{ij}$. A tripled unit cell (yellow triangles) with sublattices $A,B,C$ and nearest-neighbor vectors $\b{e}_{1,2,3}$ (orange arrows) is necessary to describe the sequence of
  field-induced ordered states in (b), 
  which comprise the coplanar Y, 2:1 canted (V), and up-up-down (UUD) states as well as 
  the non-coplanar umbrella state. In the classical limit, the coplanar and non-coplanar states
  are degenerate at all fields.}
  \label{lattice}
\end{figure}

At finite field, the QSL remains stable up to a $\kappa$-dependent critical magnetic field $h_c(\kappa)$ above which the spinon gap (\ref{gap_largen}) closes and spinons condense [Fig.~\ref{fig:gap}(b)]. As for the zero-field 120$^\circ$ state, the nature of the resulting magnetic order is encoded in the structure of the condensate amplitudes $z_{A\alpha},z_{B\alpha},z_{C\alpha}$. For efficient numerics at $h>h_c(\kappa)$, we use ans\"atze for those amplitudes that correspond to the classical ground states of the TLHAF in a magnetic field~\cite{chubukov91}. For a fixed field value below the saturation field, these form a degenerate set of states with three-sublattice order. Within this set, two classes can be further delineated that describe coplanar and non-coplanar ordering, respectively [Fig.~\ref{lattice}(b)]. Coplanar states include the Y state and the V state; the collinear up-up-down (UUD) state occurs as a special case of the Y state, for a specific field value. The non-coplanar state is the umbrella state.

To describe spinon condensation from the homogeneous $\mathbb Z_2$ QSL to the field-induced
three-sublattice order, we introduce a six-component Nambu spinor basis, 
\begin{align}\label{nambu}
\Phi_{a}(\k) = \left(b_{\k Aa\uparrow},b_{\k Ba\uparrow},b_{\k Ca\uparrow},
b^\dagger_{-\k Aa\downarrow},b^\dagger_{-\k Ba\downarrow},
b^\dagger_{-\k Ca\downarrow}\right)^T,    
\end{align}
containing the spinon degrees of freedom on the three sublattices $A,B$, and $C$, in the momentum space of the underlying triangular Bravais lattice (Appendix~\ref{app:3sub}). The effective Hamiltonian (\ref{Heff}) now reads:
\begin{align}
  \mathcal{H}_{\rm MF}
  &= N \N \frac{J}{2} \sum_{r<s} 
  |Q_{rs}|^2\nn\\ 
  &\phantom{=}+ \sum_{\k }\Phi^\dagger_{a}(\k) \Big(\hat H(\k) 
  - h \frac{\kappa}{4} \Sigma \Big)
   \Phi_{a}(\k)\lb
   &\phantom{=}- \frac{N \N}{3} (\kappa+1) \sum_r \mu_r 
   -N\N\frac{h\kappa}{4},
\label{hmf}
\end{align}
where $r,s=A,B,C$ are sublattice indices. We also define the diagonal matrix
\begin{align}
    \Sigma=\mathrm{diag}\begin{pmatrix}1,1,1,-1,-1,-1\end{pmatrix},
\end{align}
and the dynamical matrix $\hat H(\k)$ is given in Appendix~\ref{app:3sub}.

The dynamical piece of $\mathcal{H}_{\rm MF}$ can be solved by a bosonic Bogoliubov transformation, i.e., a $6\times 6$ matrix $M(\b{k})$ that obeys the pseudo-unitary condition $M(\k)^\dagger\Sigma M(\k) =\Sigma$ and rotates the Nambu spinor (\ref{nambu}) according to $\Phi_{a}(\k)=M(\b{k})\Upsilon_{a}(\k)$ where
\begin{align}\label{Upsilon}
    \Upsilon_{a}(\k) &= 
  \bigl(\xi_{1a\uparrow}(\k),\xi_{2a\uparrow}(\k),\xi_{3a\uparrow}(\k),\nn\\
    &\hspace{10mm}\xi_{1a\downarrow}^\dagger(-\k),\xi_{2a\downarrow}^\dagger(-\k),\xi_{3a\downarrow}^\dagger(-\k)\bigr)^T.
\end{align}
This diagonalizes the dynamical matrix $\hat{H}(\b{k})$,
\begin{align}\label{bogoliubov}
    M^\dagger(\k)\hat H(\k) M(\k) = \hat\Omega(\k),
\end{align}
where
\begin{align}
    \hat{\Omega}(\b{k})&=\mathrm{diag}\bigl(E_1(\k),E_2(\k),E_3(\k),\nn\\
    &\hspace{15mm}E_1(-\k),E_2(-\k),E_3(-\k)\bigr),
\end{align}
contains the eigenvalues $E_n(\b{k})$ which depend implicitly on the bond order parameters $Q_{rs}$ and Lagrange multipliers $\mu_r$, and are obtained numerically following the algorithm in Ref.~\cite{wessel05}. Here $n=1,2,3$ is a band index, and $\xi_{na\sigma}(\b{k}),\xi^\dag_{na\sigma}(\b{k})$ in (\ref{Upsilon}) are the corresponding Bogoliubov eigenoperators. The dispersing spinons have the spin-split spectrum $\omega_{n\pm}(\k) = E_{n}(\k) \pm \kappa |h|/4$. Field-induced magnetic order ensues when the lowest-lying Bogoliubov eigenmode $\omega_{n-}(\b{k})$ touches zero at the critical field $h=h_c(\kappa)$; this spin-gap closing is found to always occur at $\b{k}=\b{0}$ in the reduced (three-sublattice) Brillouin zone and signals the onset of a spinon BEC.

The spinon BEC for $h>h_c(\kappa)$ is described by $\langle\Phi_{a}(\b{0})\rangle = \sqrt{(\mathcal N/3)} Z$ where the complex vector
\begin{align}
Z \equiv (z_{A\uparrow},z_{B\uparrow},z_{C\uparrow},
z^*_{A\downarrow},z^*_{B\downarrow},
z^*_{C\downarrow})^T
\end{align}
of expectation values describes three-sublattice classical ordering
via the map 
\begin{align}\label{SpinOrder}
\langle\b{S}_r\rangle =\frac{1}{2}z_{r\alpha}^*\b{\sigma}_{\alpha\beta}z_{r\beta},\hspace{5mm}r=A,B,C.
\end{align}
In the BEC phase, the ground-state energy density $\varepsilon_0$ has contributions from both condensed ($\b{k}=\b{0}$) and uncondensed ($\b{k}\neq\b{0}$) spinons:
\begin{align}
  \frac{\varepsilon_0}{N}& = \frac{J}{2} \sum_{r<s}\left(|Q_{rs}|^2+\frac{\kappa^2}{2}\right) \lb
   &\phantom{=}+ \frac{1}{3}Z^*(h) 
   \left[\hat H(\b{0}) - \left(h \frac{\kappa}{4}+\frac{B_z}{2}\right) \Sigma \right]Z(h)\lb
   &\phantom{=}+ \frac{1}{\N} \sum_n\sum_{\k\neq \b{0}}
   E_{n}(\b{k})
   - \frac{1}{3} (\kappa+1) \sum_r \mu_r.
\label{emfcond}
\end{align}
The large-$N$ saddle-point equations (\ref{eom_master}) reduce to minimizing the ground-state energy density $\partial\varepsilon_0/\partial Q_{rs}^*=0$, $\partial\varepsilon_0/\partial\mu_r=0$ and can be written explicitly as:
\begin{align}
    Q_{rs}&=\epsilon_{\alpha\beta}z_{r\alpha}z_{s\beta}-\frac{2}{\N J}
    \sum_n\sum_{\k\neq \b{0}}\frac{\partial E_{n}(\k)}{\partial Q^*_{rs}},\label{MF1}\\
    \kappa&=|z_{r\alpha}|^2-1+\frac{3}{\N}\sum_n\sum_{\k\neq \b{0}}\frac{\partial E_{n}(\k)}{\partial\mu_r}.\label{MF2}
\end{align}
In a BEC phase, the third saddle-point equation $\delta\mathcal S/\delta b_{ia\alpha}^*=0$ reduces to $\partial\varepsilon_0/\partial z_{r\alpha}^*=0$ and becomes:
\begin{align}
    0&=\left(\hat H(\b{0})-(h\kappa/4)\Sigma\right)Z.\label{MF3}
\end{align}
It stipulates that the vector $Z$ of condensate amplitudes must be a zero eigenvector of the Bogoliubov problem solved earlier~\cite{sachdev92}, which specifies the ordering pattern via Eq.~(\ref{SpinOrder}). Once Eqs.~(\ref{MF1}-\ref{MF3}) have been solved self-consistently, one can substitute the mean-field solution $\{Q_{rs},z_{r\alpha},\mu_r\}$ back into Eq.~(\ref{emfcond}) to compute the ground-state energy density, and also calculate observables such as the uniform magnetization density $m$,
\begin{equation}
  m=-\left.\frac{1}{N\N}\frac{\partial\varepsilon_0}{\partial B_z}\right|_{B_z=0}.
\label{magOP}
\end{equation}

\subsection{The classical limit: $\kappa\rightarrow\infty$}
\label{sec:classical}

The coupled system of equations (\ref{MF1}-\ref{MF3}) 
cannot be solved analytically. However,
in the classical limit $\kappa\rightarrow\infty$, the contributions from uncondensed bosons become
subleading and the saddle-point equations simplify to:
\begin{align}
Q_{rs}& =\epsilon_{\alpha\beta}z_{r\alpha}z_{r\beta},\\
\kappa&=|z_{r\alpha}|^2,\\ 
 0&=\left(\hat H(\b{0})-(h\kappa/4)\Sigma\right)Z.
\end{align}
Upon substituting those equations into the mean-field energy density (\ref{emfcond}) and taking the $\kappa\rightarrow\infty$ limit (Appendix~\ref{app:classical}), Eq.~\eqref{emfcond} reduces to the classical TLHAF in a field with spin vectors given by (\ref{SpinOrder}). Thus, we consider a family of condensate solutions 
\begin{align}\label{CondensSol}
  \begin{pmatrix}
    z^*_{r\uparrow}(\theta,\phi)\\
    z^*_{r\downarrow}(\theta,\phi)
  \end{pmatrix}
  =
  e^{i\phi\sigma^z/2}
  e^{i\theta\sigma^y/2}
  \begin{pmatrix}
    \sqrt{n_c}\\
    0
  \end{pmatrix}
  e^{-i\theta\sigma^y/2}
  e^{-i\phi\sigma^z/2},
\end{align}
parameterized by angles $\theta(h),\phi(h)$ and the condensate magnitude $n_c$, such that the spin arrangements associated with these solutions give rise to the three-sublattice classical ground states of the TLHAF in the presence of a magnetic field~\cite{chubukov91}.
Two particular classes of these solutions, that we term $Z_{\rm cop.}(h)$ and $Z_{\rm noncop.}(h)$ respectively,
describe the coplanar and non-coplanar ordered states at various magnetic field magnitudes $h$ [Fig.~\ref{lattice}(b)]. In the classical TLHAF, the Y state stabilized for $h<3J$ becomes the collinear UUD state at $h=3J$, which tilts into the V state 
for $3J<h<9J$, and eventually attains the fully polarized state for $h> h_\text{FP} = 9J$. The UUD state has a net magnetization given by $1/3$ of the saturation value in the fully polarized state. For each field value $0< h< 9J$, the corresponding coplanar state is degenerate with a non-coplanar umbrella state. In Appendix~\ref{app:ansatze}, we provide the explicit expressions for those various condensate 
wave functions.

\subsection{Finite $\kappa$: numerical solution of the mean-field equations}

For finite $\kappa$, the mean-field equations (\ref{MF1}-\ref{MF3}) 
are 
solved numerically. Guided by the phenomenology of the rare-earth delafossites, we conjecture that the magnetic orders of the $\kappa\rightarrow\infty$ limit persist at finite $\kappa$ but with shifted values of the various critical fields, and renormalized values of the magnetic moment and ground-state energy densities due to quantum corrections. In particular, the condensate amplitude $n_c$ in the mean-field ansatz (\ref{CondensSol}), which is akin to a staggered moment in our three-sublattice ordered states, is renormalized by quantum fluctuations according to:
\begin{align}
n_c &= \frac{1}{3}\sum_r
|z_{r\alpha}|^2\nn\\
&= \kappa+1-\frac{1}{\N}\sum_r\sum_n\sum_{\k\neq\b{0}}
\frac{\partial E_{n}(\k)}{\partial\mu_r}.
\label{stagMom}
\end{align}
In practice, we seed the mean-field equations with our classical mean-field ans\"atze, iterate those equations until convergence is reached, and compare the ground-state energy densities of the various solutions (local minima) to determine the global minimum (Appendix~\ref{app:ansatze}).

\begin{figure}
  \subfloat[]{
    \includegraphics[width=0.47\columnwidth]{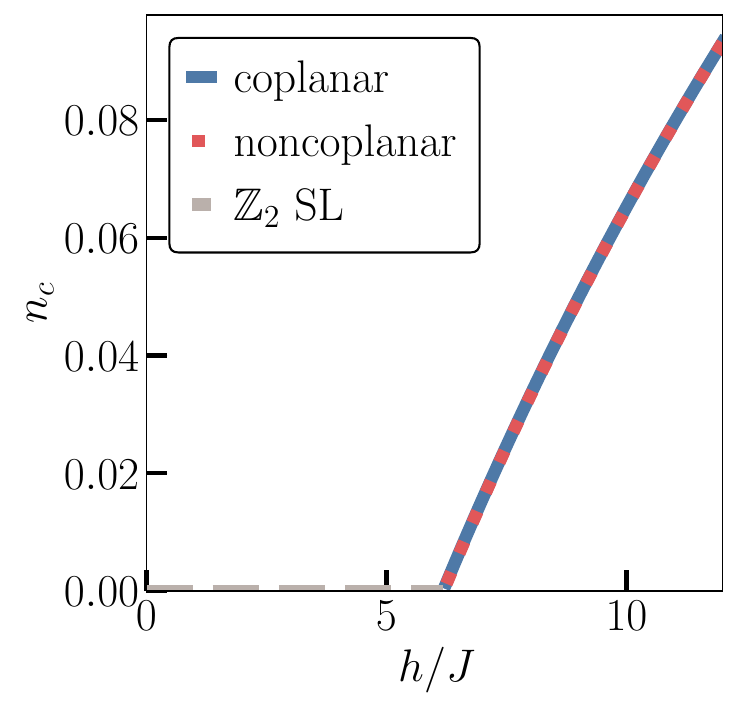}
  }
  \hfill
  \subfloat[]{
    \includegraphics[width=0.47\columnwidth]{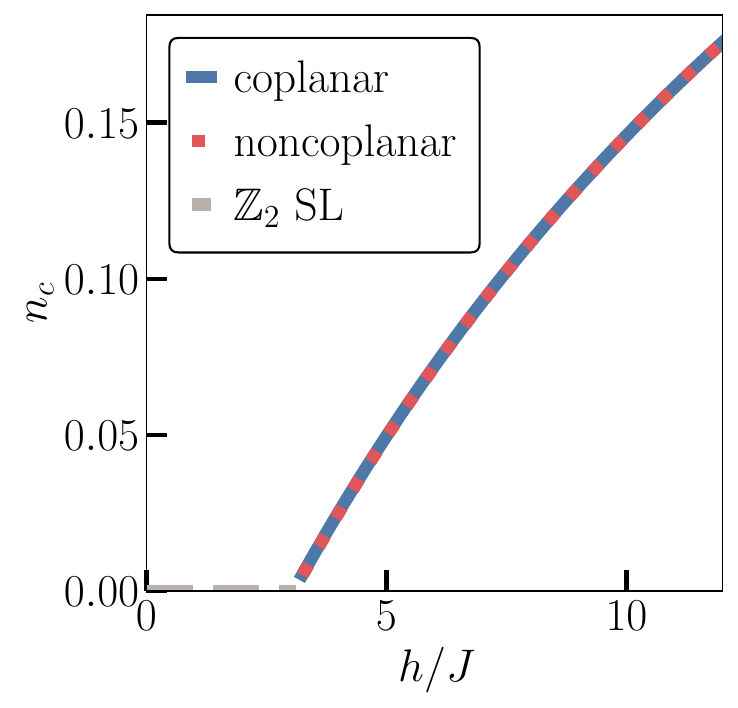}
  }\\
  \subfloat[]{
    \includegraphics[width=0.47\columnwidth]{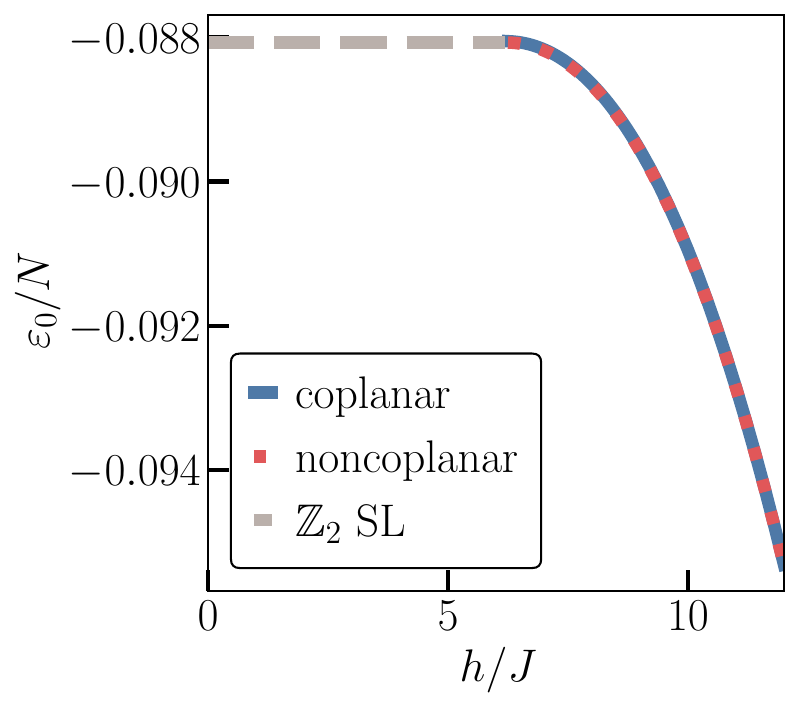}
  }
  \hfill
  \subfloat[]{
    \includegraphics[width=0.47\columnwidth]{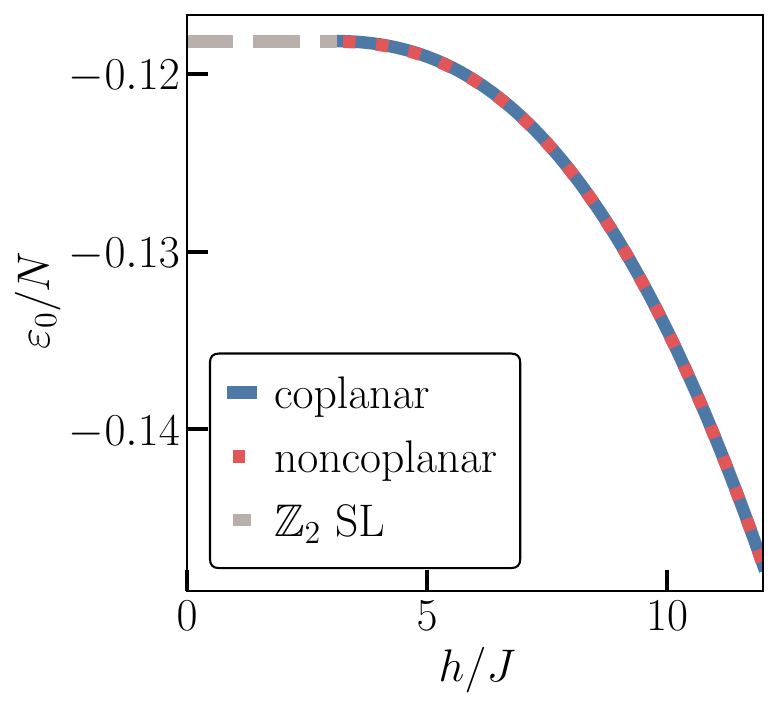}
  }
  \caption{Transition from $\mathbb{Z}_2$ QSL to field-induced order in the large-$N$ limit for (a,c) $\kappa=0.18$ and (b,d) $\kappa=0.24$. Panels (a) and (b) show the magnitude $n_c$ of the spinon BEC (three-sublattice staggered moment) rising continuously from the QSL following the onset of coplanar (blue) or non-coplanar (orange) order beyond the critical fields $h_c(\kappa)$. Panels (c) and (d) compare the ground-state energy densities of those two competing states; modulo numerical accuracy, they are degenerate as in the classical limit. 
  }
\label{largeN_a}
\end{figure}

The numerical solution of the mean-field equations reveals a field-induced second-order transition between the $\mathbb Z_2$ QSL and three-sublattice magnetically ordered phases described by a spinon BEC. The second-order nature of the transition can be seen in Figures~\ref{largeN_a}(a) and ~\ref{largeN_a}(b) where the condensate magnitude $n_c$ is seen to increase continuously from the QSL phase as the external magnetic field is increased above its critical value $h_c(\kappa)$. However, in this strict large-$N$ limit, we find that the coplanar and non-coplanar solutions remain degenerate modulo numerical accuracy for finite $\kappa$ 
[Figures~\ref{largeN_a}(c) and ~\ref{largeN_a}.(d)], as in the classical $\kappa\rightarrow\infty$ limit. Although the large-$N$ energy density (\ref{emfcond}) includes quantum zero-point fluctuations via the contribution of uncondensed spinons, these quantum fluctuations do not conclusively lift the degeneracy between the competing coplanar and non-coplanar states.

\begin{figure}
  \subfloat[]{
    \includegraphics[width=0.47\columnwidth]{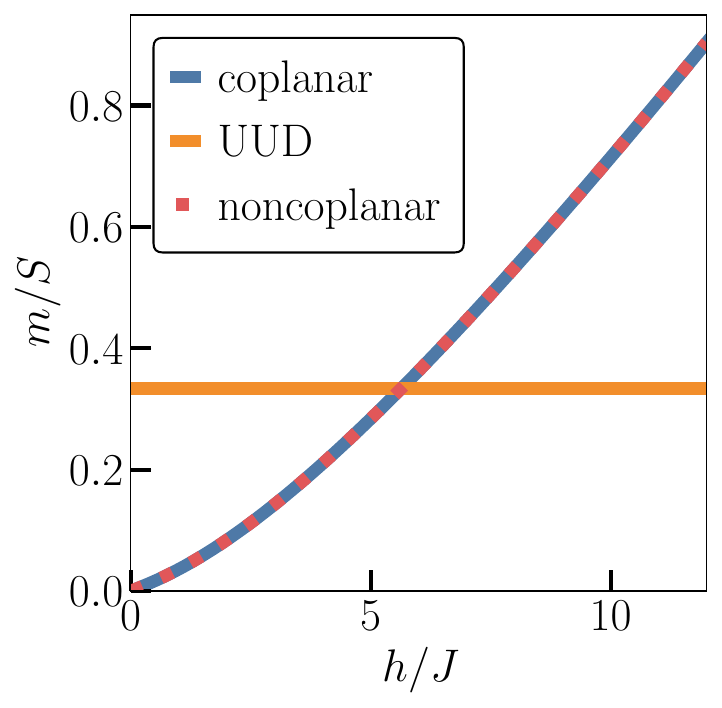}
  }
  \hfill
  \subfloat[]{
    \includegraphics[width=0.47\columnwidth]{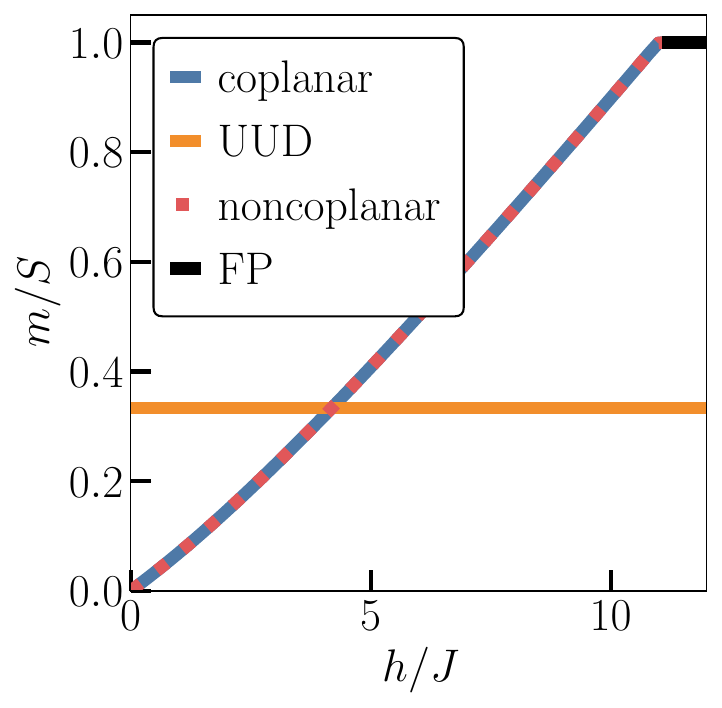}
  }\\
  \subfloat[]{
    \includegraphics[width=0.47\columnwidth]{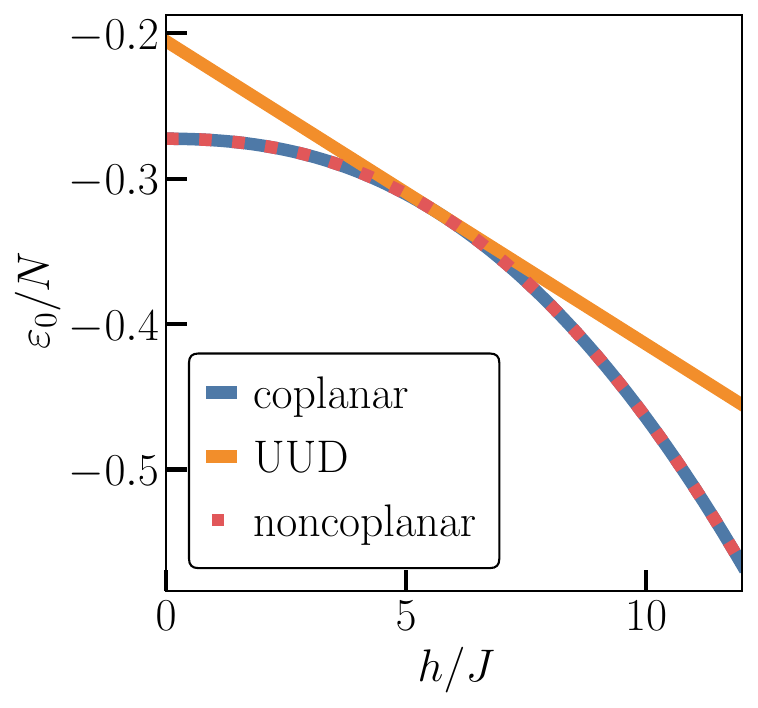}
  }
  \hfill
  \subfloat[]{
    \includegraphics[width=0.47\columnwidth]{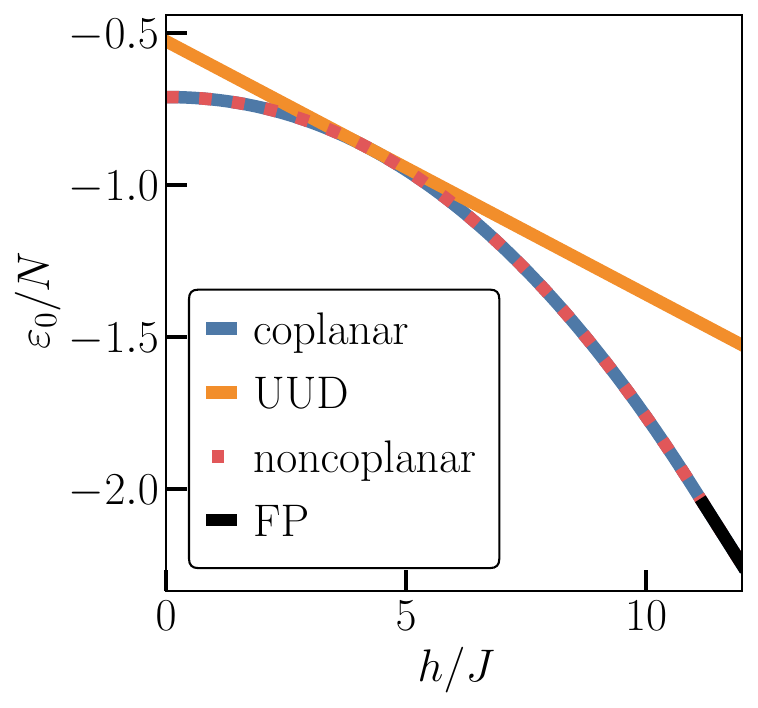}
  }
  \caption{Comparison of (a,b) the uniform magnetization $m$ [Eq.~(\ref{magOP})] and (c,d) the energy density for various solutions of the large-$N$ mean-field equations: the coplanar state (blue), 
  the high-energy up-up-down state (orange) 
  and the non-coplanar state (red). Values of $\kappa$ for (a,c) is $\kappa=0.5$ and for (b,d) is $\kappa=1$. 
  The UUD solution has generally higher energy but becomes degenerate with the non-collinear/non-coplanar solutions at a discrete value of $h$, as in the classical limit.}
\label{largeN_b}
\end{figure}

To investigate the possibility of a 1/3 magnetization plateau, we also consider the fate of the collinear UUD 
state, treated as a variational ansatz for all fields $h\geq h_c(\kappa)$ [Fig.~(\ref{largeN_b})]. 
As mentioned before, in the classical limit, the UUD state is stabilized at a single field value $h=3J$. 
Here, we find that the UUD ansatz yields a solution of the large-$N$ mean-field equations for a wide range of fields 
(see Fig.~\ref{largeN_b}).
The lower symmetry of the UUD spin configuration allows us to identify its associated variational
wave function both numerically and analytically. As a function of the external magnetic field,
both the coplanar and noncoplanar mean-field 
solution branches are continuously connected to the sublattice-symmetric
$120^\circ$ ground state, and both of them 
attain homogeneous (but different) local chemical potentials 
for all field values.
The optimal 
energy UUD solution, 
on the other hand, breaks the sublattice occupational symmetry in favor of a configuration 
with $\mu_A=\mu_B\neq \mu_C$.
Its energy density is, however,
higher than that of the non-collinear (coplanar) and non-coplanar states, except at isolated field values 
which differ from the classical value $h=3J$ [Figures~\ref{largeN_b}(b) and ~\ref{largeN_b}(c)].

Finally, at saturation fields higher than the classical value $h=9J$, the fully
polarized state sets in continuously for all $\kappa$. Figure~\ref{largeN_b}(b)
shows the onset for the polarized state for $\kappa=1$. The critical field
for the polarization transition decreases towards
the classical value with increasing $\kappa$.

Overall, we conclude that the large-$N$ limit of the Schwinger-boson theory of the quantum TLHAF in a magnetic field is able to capture a continuous field-induced quantum phase transition from a $\mathbb{Z}_2$ QSL to ordered states with net magnetization. However, despite accounting for some measure of quantum fluctuations, the large-$N$ limit does not resolve a degeneracy between coplanar and non-coplanar states that is also encountered in the classical TLHAF. Therefore, the order-by-disorder mechanism present in the solution of the TLHAF via the $1/S$ expansion~\cite{chubukov91} is not captured by Schwinger-boson mean-field theory. We next show that the inclusion of $1/N$ corrections lifts the large-$N$ degeneracy and also predicts a 1/3 magnetization plateau with UUD order, in agreement with the phenomenology of the rare-earth delafossites.

\section{$1/N$ corrections}
\label{fluctuation}

Beyond the strict large-$N$ (mean-field) limit, $SU(N)$ and Sp(N) Schwinger-boson theories admit controlled $1/N$ expansions~\cite{arovas88,read91,sachdev91,sachdev92,auerbach94}. The $\mathbb{Z}_2$ QSL is a gapped phase with a discrete dynamical gauge field, and $1/N$ corrections do not substantially alter its physics. However, as has been pointed out recently, $1/N$ corrections are crucial to capture even qualitative aspects of ordered phases in the Schwinger-boson formalism~\cite{zhang19,zhang22}. In principle, all observables computed in Schwinger-boson theory, such as the ground-state energy density $\varepsilon_0$ and the uniform magnetization $m$, receive fluctuation corrections suppressed by powers of $1/N$,
\begin{align}
\frac{\varepsilon_0}{N}&=\varepsilon_0^{(0)}+\frac{\varepsilon_0^{(1)}}{N}+\frac{\varepsilon_0^{(2)}}{N^2}+\dots,\label{corr-ener}\\
m&=m^{(0)}+\frac{m^{(1)}}{N}+\frac{m^{(2)}}{N^2}+\dots,\label{corr-m0}
\end{align}
where $\varepsilon_0^{(0)}$ and $m^{(0)}$ correspond to the large-$N$ values computed in Sec.~\ref{largeN}. In broken-symmetry phases, these corrections can become substantial~\cite{zhang22}, in particular in the case of SU(2) spins with $N=1$. Here, we only consider the first (subleading) correction, to order $1/N$. After reviewing the $1/N$ expansion and its diagrammatic formulation (Sec.~\ref{sec:1/N}), we show that $1/N$ corrections becomes vital in capturing the subtle order-by-disorder mechanism in the TLHAF model (Sec.~\ref{sec:ObD}), and also in reproducing the correct magnon excitation spectrum in the ordered phases (Sec.~\ref{sec:Sqw}).

\subsection{The $1/N$ expansion}
\label{sec:1/N}

In the $N\rightarrow\infty$ limit, the saddle-point evaluation of the Schwinger-boson partition 
function [Eq.~\eqref{actPart}] becomes exact. In this limit, the
bond order parameters $Q_{ij}$, Lagrange multipliers $\lambda_i$, 
and condensate amplitudes $z_{ia\alpha}$ 
are all determined self-consistently. 
At finite $N$, these quantities acquire fluctuations. Fluctuations of $\lambda_i$ are akin 
to a dynamical temporal gauge field which imposes the local constraint (\ref{constraint}) 
on the Schwinger-boson density. This leads to (gauge) zero modes in the inverse fluctuation 
propagator and in turn, unphysical divergences in the partition function, which can be avoided 
by discarding such fluctuations \cite{ghioldi18,trumper97}. 
Thus, we only consider fluctuations in the bond
parameters, $A_I(\b{R})\in\{\delta Q_{ij},\delta Q_{ij}^*\}$
(here $(\b{R},r)=i$ is the unit-cell, three-sublattice index for site $i$, 
and $I$ denotes the $9$ independent bonds 
$(ij)$ 
associated to
each unit cell; see Appendix~\ref{vertex} for 
more details on this notation), which leads to the effective
action:
\begin{align}
    \c{W}=\c{W}_\text{mf}+\c{W}_\text{fluc},
\end{align}
where
\begin{align}
    \c{W}_\text{mf}=\N\biggl[&\frac{J}{2}\sum_{r<s}\left(|Q_{rs}|^2+\frac{\kappa^2}{2}\right)\nn\\
    &-\frac{1}{3}(\kappa+1)\sum_r\mu_r
    -\left(\frac{h\kappa}{4}+\frac{B_z}{2}\right)\biggr]\nn\\
    &+\frac{1}{N}\Tr\ln \hat G_{\Phi}^{-1}(Q,Q^*,\lambda)
    ,
\end{align}
is the leading-order (mean-field) action with a trace
over the inverse Schwinger-boson Green's function obtained for the saddle-point fields $Q_{rs}$ and $\mu_r$, and,
\begin{align}
    \c{W}_\text{fluc}
    =&
    \frac{1}{2}\int^\beta_ 0 d\tau\sum_{\b{R}}A_I(\tau,\b{R}) [D_{A}^{-1}]^{IJ}
    A_J(\tau,\b{R})\nn\\
    &+
    \frac{1}{N}\Tr\ln\left(1+\hat G_\Phi(Q,Q^*,\lambda)\hat A_I\hat\Gamma^I\right)
    ,
\end{align}
is the subleading contribution from the bond fluctuations. For the $1/N$ corrections to the observables, it
suffices to consider the leading-order expansion 
term of the fluctuation action that gives rise to a
three-point interaction vertex $\Gamma^I$ between the Schwinger 
bosons and the bond fluctuations. 

\begin{figure}
    \includegraphics[width=0.95\columnwidth]{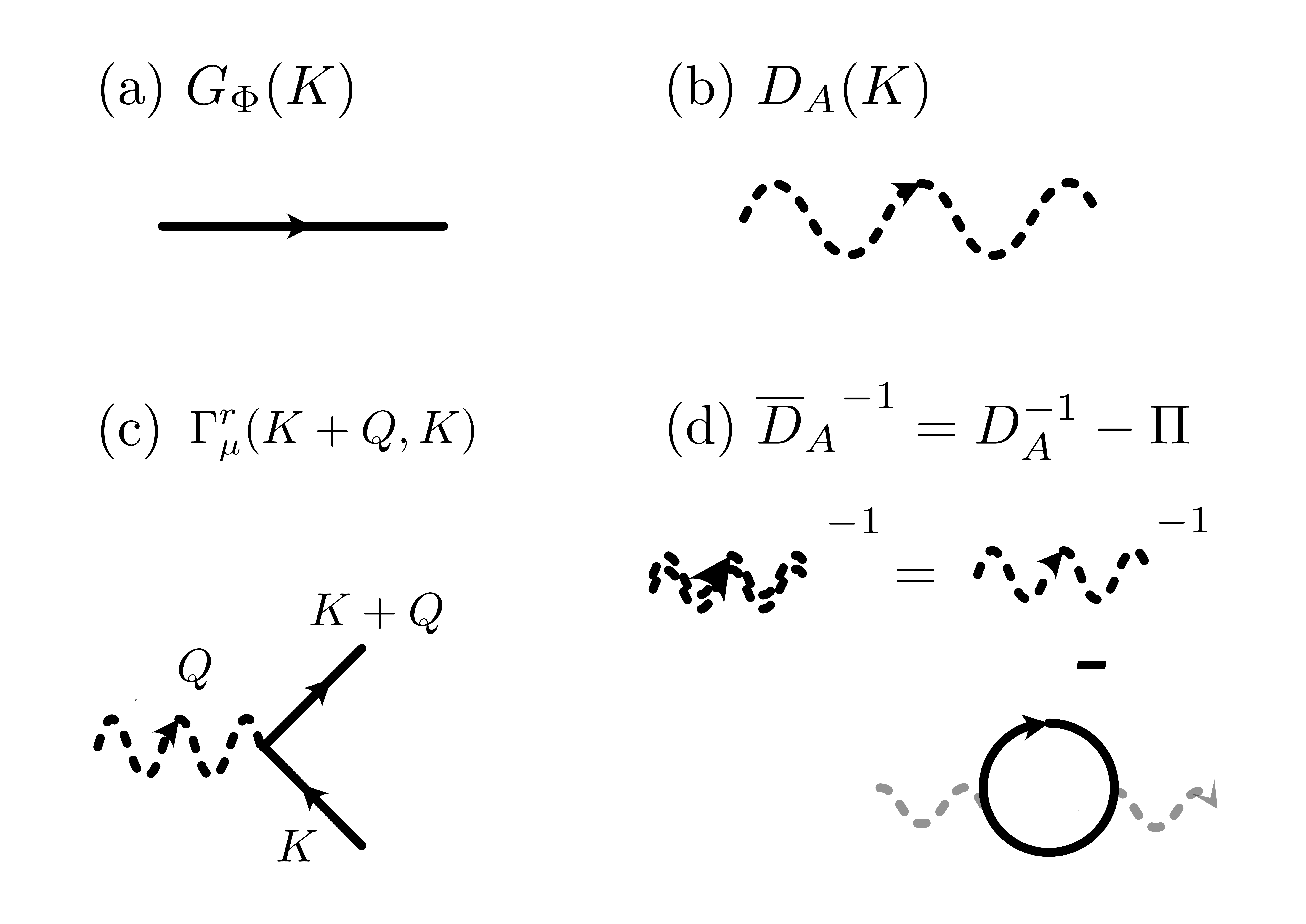}
  \caption{Feynman rules for the $1/N$ expansion: (a) Schwinger-boson propagator; (b) bond-order-parameter fluctuation propagator; (c) three-point interaction. $1/N$ corrections involve the RPA propagator $\overline{D}_A$ (d), obtained from the bare propagator $D_A$ and the one-loop irreducible self-energy $\Pi$ (polarization bubble). 
  }
\label{vacc_pol}
\end{figure}

These various contributions to the action can be represented diagrammatically in
the frequency-momentum space $K=(-ik_0,\b{k})$ (Fig.~\ref{vacc_pol}). 
The non-interacting part of the action functional 
involves the bare propagators $[G_\Phi(K)]_{ab}=
\langle a,K|\hat G_{\Phi}|b,K\rangle$ and $D_A(K)$ for Schwinger bosons 
and bond order-parameter fluctuations, respectively; they are expressed as
\begin{align}
  [G_{\Phi}(K)]_{ab} &= \delta_{ab}
  \left[-\left(ik_0+\frac{h\kappa}{4}+\frac{B_z}{2}\right)\Sigma
+\hat H(\k)\right]^{-1},\label{bosonGF}\\
  D_{A}(K) &= \frac{1}{\N}\begin{bmatrix}
    0&
    \frac{1}{2}\hat{I}
    \\
    \frac{1}{2}\hat{I} &0
  \end{bmatrix},  \text{ where } \hat I_{IJ} =\delta_{IJ},
\end{align}
which we denote by solid and wiggly lines, respectively [Figures~\ref{vacc_pol}(a) and \ref{vacc_pol}(b)]. 
The $1/N$ contribution from $\c{W}_\text{fluc}$ involves the three-point vertex 
$\langle K+P|\hat\Gamma^I|K\rangle = \Gamma^I(K+P,K)$ 
which corresponds to the decay of a bond-order-parameter fluctuation into a pair of 
Schwinger bosons [Fig.~\ref{vacc_pol}(c)]. The explicit expression for $\Gamma^I$ is cumbersome 
and therefore relegated to Appendix~\ref{vertex}.

\subsection{Order by disorder from $1/N$ corrections: coplanar order and a magnetization plateau}
\label{sec:ObD}

The first effect of $1/N$ corrections is to correct the ground-state energy density. The $1/N$ correction $\varepsilon_0^{(1)}$ in Eq.~(\ref{corr-ener}) is given by~\cite{arovas88}:
\begin{align}\label{ener-1/N}
    \varepsilon_0^{(1)}=\frac{1}{2\beta\N}\sum_P\tr\ln\overline{D}_A^{-1}(P),
\end{align}
where
\begin{align}
    \overline{D}_A^{-1}(P)=D_A^{-1}(P)-\Pi(P),
    \label{DRPA}
\end{align}
is the RPA-resummed fluctuation propagator [Fig.~\ref{vacc_pol}(d)], and
\begin{align}
\Pi^{IJ}(P)=\frac{3}{\N}\sum_{K}&\tr\bigl(
  \Gamma^I(K+P,K)
  G_{\Phi}(K)
  \nn\\
    &\times \Gamma^J(K,K+P)G_{\Phi}(K+P)\bigr),
\end{align}
is the one-loop fluctuation self-energy (polarization bubble) obtained by restricting the momentum
sum to the first Brillouin zone. 
Traces ($\tr$) in these 
expressions are over the appropriate index space (Schwinger-boson Nambu space or fluctuation-field index space, respectively) and
the factor of $3$ comes from the three-sublattice reduction of the Brillouin zone. Details of the fluctuation self-energy
calculation are provided in Appendix~\ref{fluctSelf}.

\begin{figure}[t]
  \subfloat[]{
    \includegraphics[width=0.49\columnwidth]{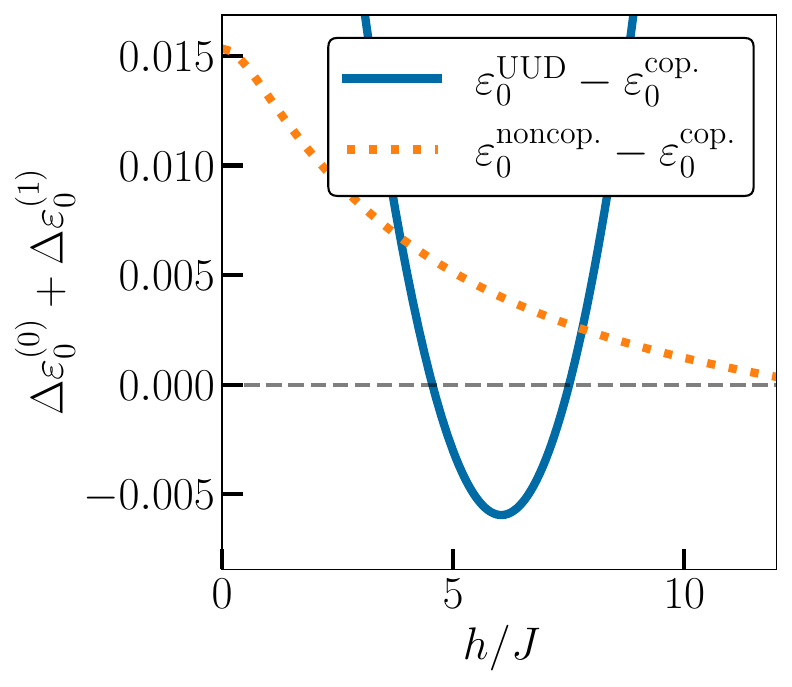}
  }
  \subfloat[]{
    \includegraphics[width=0.49\columnwidth]{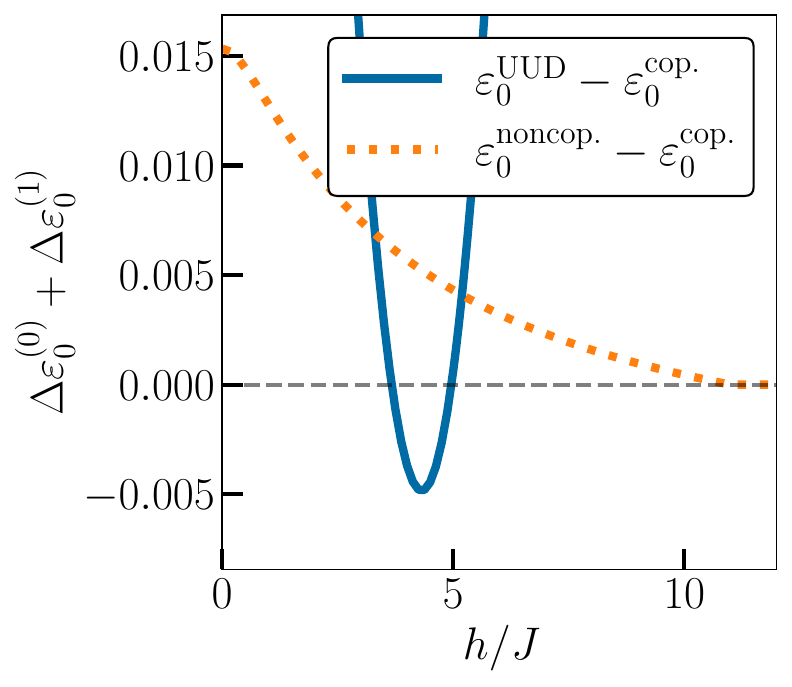}
  }
  \caption{The differences between the 
  $1/N$-corrected ground-state energy densities of the coplanar and non-coplanar (orange), and coplanar 
  and UUD states (blue), extrapolated to the SU(2) case ($N=1$), for 
  (a) $\kappa=1/2$ and (b) $\kappa=1$. $1/N$ corrections favor the coplanar states 
  and also stabilize the UUD state in a finite field range.
  }
  \label{fluctN_a}
\end{figure}

\begin{figure}[t]
  \subfloat[]{
    \includegraphics[width=0.47\columnwidth]{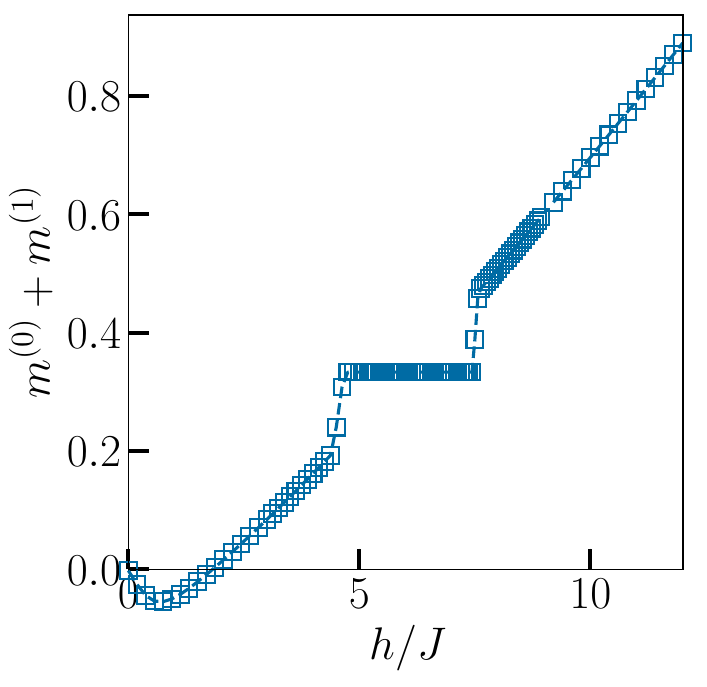}
  }
  \hfill
  \subfloat[]{
    \includegraphics[width=0.47\columnwidth]{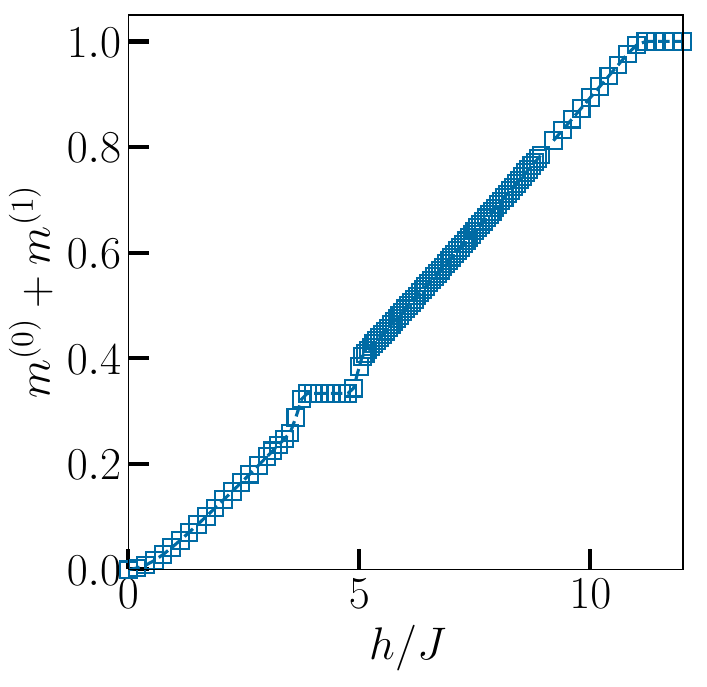}
  }
  \caption{$1/N$-corrected uniform magnetization $m$ extrapolated to the SU(2) limit ($N=1$) for 
    (a) $\kappa=1/2$, (b) $\kappa=1$ with a finite-sized Brillouin zone of $36\times 36\times 3$ 
    lattice sites.
    The UUD state stabilized by $1/N$ corrections (Fig.~\ref{fluctN_a}) results in a 1/3 magnetization plateau 
    in an intermediate field range.
    For smaller fields, the correction introduces anomalous diamagnetic behavior that 
    diminishes with increasing system size. 
  }
\label{fluctN_b}
\end{figure}

In Fig.~\ref{fluctN_a}, we plot the $1/N$-corrected energy density $\varepsilon_0/N\approx\varepsilon_0^{(0)}+\varepsilon_0^{(1)}/N$, extrapolated to the physical $N=1$ limit of SU(2) spins. Remarkably, the $1/N$ correction lifts the degeneracy between the large-$N$ coplanar and non-coplanar solutions in favor of coplanar order, for all ranges of $\kappa$ and $h$ for which ordered phases win over the $\mathbb{Z}_2$ QSL. Thus, the $1/N$-corrected Schwinger-boson theory is able to capture the quantum order-by-disorder mechanism which stabilizes coplanar states in the semiclassical ($1/S$) expansion of the TLHAF in a magnetic field~\cite{chubukov91}. Furthermore, for a select range of applied magnetic fields, the collinear UUD state becomes the state of lowest energy. This results in a 1/3 magnetization plateau for all $\kappa$ outside the QSL state. Figure~\ref{fluctN_b} plots the $1/N$-corrected uniform magnetization $m\approx m^{(0)}+m^{(1)}/N$, obtained from the probe-field derivative of the $1/N$-corrected energy density [Eq.~(\ref{corr-m0})], extrapolated to the SU(2) limit $N=1$. Transitions in and out of the plateau are of first order, as in semiclassical studies~\cite{takano11,coletta16}. The width of the plateau decreases with increasing $\kappa$ and reduces to a point in the classical limit (Fig.~\ref{fluctPd}). The growth of the fluctuation contribution with 
decreasing spin-size reflects an expected behavior as the system becomes more quantum mechanical. For smaller $\kappa$ 
and applied fields, the diamagnetic first-order correction to the mean-field energy overwhelms the leading contribution
and leads to anomalous magnetization (Fig.~\ref{fluctN_b}). A similar low-field
anomalous magnetization was reported in Ref.~\cite{ghioldi18} for the triangular lattice Sp(N)
theory. The authors found that for $\kappa=1$, the anomaly disappears in the thermodynamic limit. We witness
the same outcome in Fig.~\ref{fluctN_b}~(b). However,
for the numerically accessible system sizes in our computation, the low-field anomaly
survives for smaller spin sizes. The anomaly does not affect the plateau phase as it always
appears at elevated fields with a magnetization jump $\Delta m <S/3$.

\subsection{Dynamic spin structure factor and magnon spectrum}
\label{sec:Sqw}

\begin{figure*}
  \captionsetup[subfigure]{labelfont=normalsize}
  \subfloat[]{
    \includegraphics[width=0.4\linewidth]{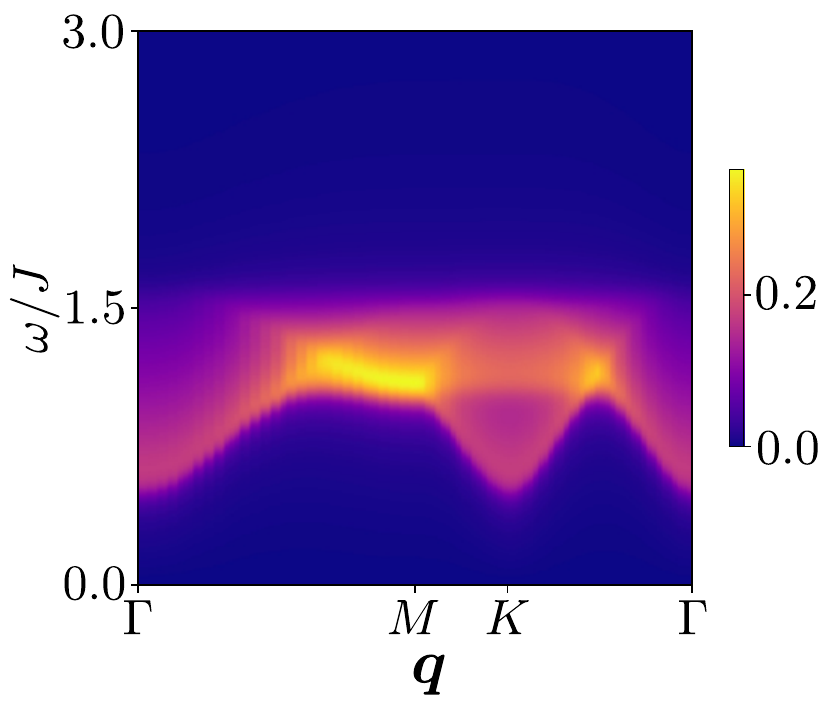}
  }
  \subfloat[]{
    \includegraphics[width=0.42\linewidth]{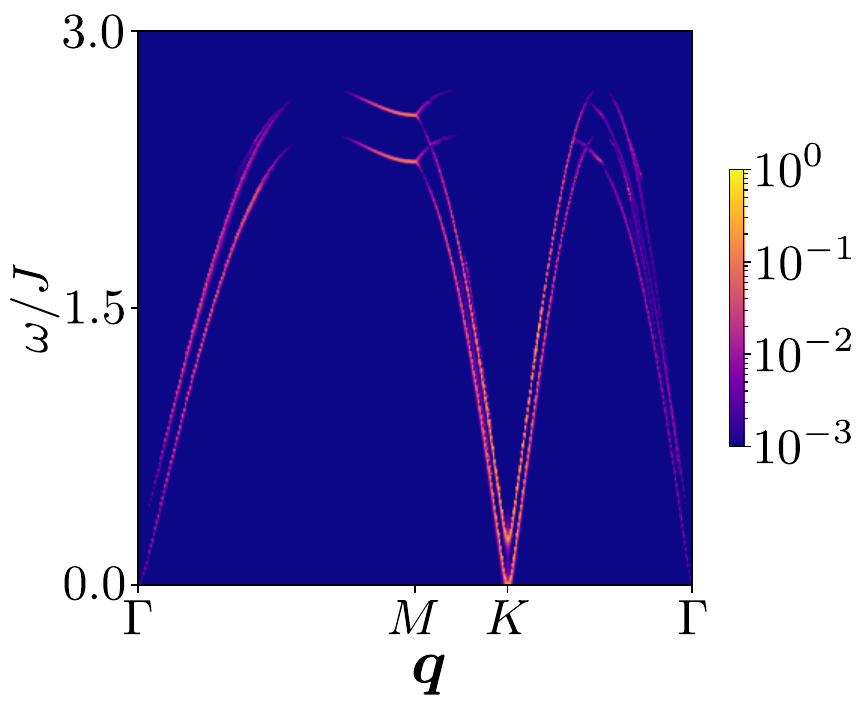}
  }\\
  \subfloat[]{
    \includegraphics[width=0.42\linewidth]{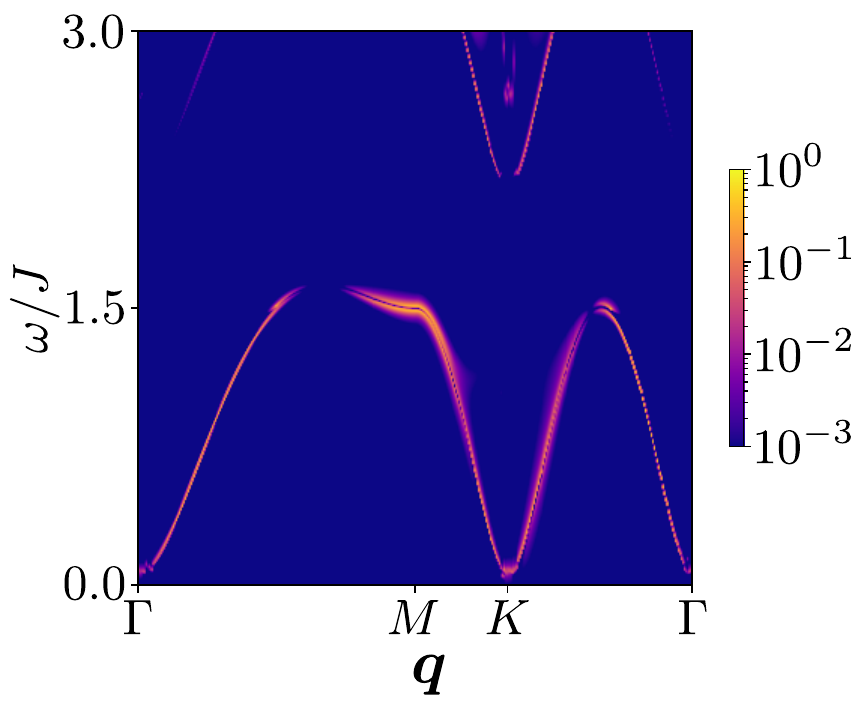}
  }
  \subfloat[]{
    \includegraphics[width=0.42\linewidth]{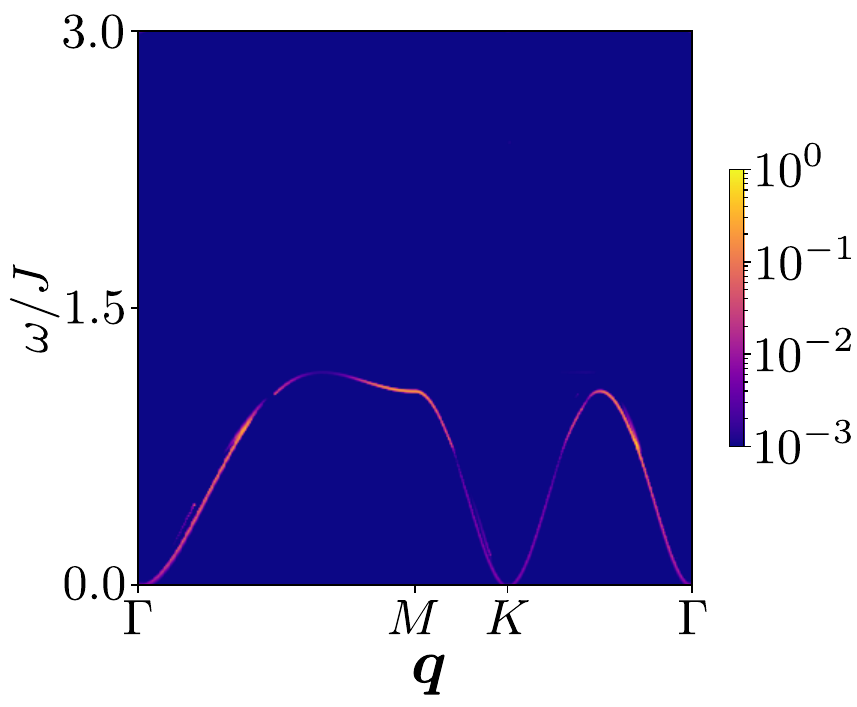}
  }
  \caption{Dynamic spin structure factor $S(\omega,\b{q})$
  along high-symmetry lines in the extended Brillouin zone, in (a) the $\mathbb{Z}_2$ QSL ($\kappa = 0.2$, $h = 0.1$), 
  (b) the Y state ($\kappa=1$, $h=0.5J$), (c) the UUD state ($\kappa=1$, $h=4.3J$), and (d) the V state ($\kappa=1$, $h=6J$). 
  These structure factors were obtained from an analytic continuation of the dynamical
  susceptibility (Appendix~\ref{SpinStruct}).}
  \label{structure}
\end{figure*}

In addition to affecting the ground-state energy of various mean-field solutions, $1/N$ corrections have been recently shown to be crucial in accounting for the correct excitation spectrum of ordered phases in the Schwinger-boson formalism~\cite{ghioldi18,zhang19,zhang22}. This manifests in computations of the dynamic spin structure factor~\cite{mourigal13},
\begin{align}\label{Sqw}
    S^{\alpha\beta}(\omega,\b{p})=\frac{1}{\N}\sum_{ij}\int dt\,e^{i\omega t-i\b{p}\cdot(\b{r}_i-\b{r}_j)}\langle S_i^{\alpha}(t)S_j^\beta(0)\rangle,
\end{align}
where $\b{r}_i$ is the position vector of site $i$. This structure factor can be obtained from an
analytic continuation of the Matsubara spin susceptibility $\chi^{\alpha\beta}(ip_0,\b{p})$ (Appendix~\ref{SpinStruct}) 
and, in an ordered phase, it should exhibit delta-function peaks at single-magnon energies. 
In the large-$N$ limit, $\chi$ is a convolution of bare Schwinger-boson propagators $G_\Phi$ (bubble diagram), which contain poles at single-spinon excitation energies
following the onset of condensation. 
For collinear (N\'eel) states on bipartite lattices, this convolution reproduces the correct magnon spectrum~\cite{auerbach1988b}, but spurious single-spinon poles remain in the large-$N$ limit for non-collinear states on frustrated lattices. As shown in References~\cite{ghioldi18,zhang19,zhang22}, inclusion of $1/N$ diagrams in the computation of $\chi$ cancels the single-spinon poles and reproduces the correct magnon peaks in $S^{\alpha\beta}(\omega,\b{p})$, as would be observed in inelastic neutron scattering. Ref.~\cite{ghioldi18} was able to demonstrate this cancellation for the 120$^\circ$ spiral order in the TLHAF at zero field.

Here, we compute these $1/N$ corrections for the various field-induced ordered phases appearing in the phase diagram of 
Fig.~\ref{fluctPd}. Details of the computation are provided in Appendix~\ref{SpinStruct}, and the resulting structure factors $S(\omega,\b{p}) = \sum_\alpha S^{\alpha\alpha}(\omega,\b{p})$ are displayed in Fig.~\ref{structure}. The emergence of the magnon signal and disappearance of the single-spinon pole in the ordered phases 
[Figures~\ref{structure}(b)-\ref{structure}(d)] are consistent with the notion that those phases are not fractionalized (spinons are confined). 
As expected from previous studies~\cite{syromyatnikov23}, the excitation spectrum is strongly affected by the external magnetic field. 
Crucially, in the collinear UUD state, a clear magnon gap arises [Fig.~\ref{structure}(c)], as opposed to the gapless 
spectrum of the coplanar Y and V states in Figures~\ref{structure}(b) and \ref{structure}(d), respectively. The character of the 
bond fluctuations is significantly different in the collinear state. The non-collinear and non-coplanar condensates
admit slow changes in the bond profiles commensurate with local spin rotations along the field-axis. Consequently,
gapless spin-wave modes arise through the $1/N$ fluctuation correction.
In the UUD configuration, such changes amount to phase twists of the bond parameters with no change
in spin texture. Thus, the unbroken  global $U(1)$ spin-rotation
symmetry elevates to a local gauge symmetry of the fluctuation action in the collinear state. With the Schwinger bosons
charged under this symmetry, the presence of the condensate triggers the Anderson-Higgs mechanism, and a gapped spectrum results. The gap signifies that the UUD state is incompressible and guarantees the stability of its magnetization plateau. For a unit-charge condensate, the Higgs phase is adiabatically connected to the confined phase~\cite{fradkin1979}, thus the UUD order is conventional.

Contrasting the sharply peaked signature of the field-ordered states, the proximate $\mathbb{Z}_2$ QSL has a diffuse 
spin structure factor generated by the two-particle continuum of its 
fractionalized quasiparticles [Fig.~\ref{structure}(a)]. Confinement in the ordered phases does not preclude a 
two-spinon continuum in their spectra as they are composite $S=1$ excitations. They do not make an appearance in the 
spectrum as their spectral weight is lower than that of the magnon modes by several orders of magnitude. 

\section{Conclusion}
\label{discussion}

Motivated by recent experimental observations of spin-liquid behavior and field-induced magnetic orders in rare-earth delafossite magnets, we have investigated the zero-temperature phase diagram of the quantum TLHAF in a magnetic field using large-$N$ Schwinger boson methods. 
For strong quantum fluctuations, tuned by a spin-size parameter $\kappa$, 
we find that a gapped $\mathbb{Z}_2$ QSL with deconfined bosonic spinons is present at small magnetic fields, but undergoes a continuous confinement transition at a critical field via spinon condensation. Beyond this critical field, the QSL gives way to a sequence of coplanar ordered states, including canted Y and V states and a 1/3 magnetization plateau with collinear up-up-down order. This sequence of field-driven phases is known from a semiclassical treatment of the TLHAF, where the classical degeneracy between coplanar and non-coplanar states is lifted in favor of the former by a quantum order-by-disorder mechanism. 
We have shown that $1/N$ corrections in the Schwinger-boson formalism have a similar effect, even away from the semiclassical limit where strong quantum fluctuations can stabilize a zero-field QSL.

Our phase diagram (Fig.~\ref{fluctPd}) qualitatively parallels the experimental observations
in various rare-earth delafossite magnets. Although the $T=0$ long-range order found here would disappear in our two-dimensional Heisenberg model at finite $T$, both weak interlayer coupling as well as spin-orbit anisotropies will restore such order at low $T$, in agreement with experiment.

We hope that our work will stimulate further theoretical and experimental studies, including the application of hydrostatic pressure and/or the exploration of other families of rare-earth delafossite materials, with the goal of completing the catalog of their phenomenology. For instance, neutron scattering and heat-transport measurements will be important to clarify the nature of the zero-field spin liquid, which in our treatment and according to some numerical studies of the $J_1$-$J_2$ model~\cite{zhuwhite15,hu2015} is a gapped $\mathbb{Z}_2$ QSL, but other such studies favor a gapless $U(1)$ QSL~\cite{iqbal16,hu2019}. 
We note that gapless spin liquids, such as a state with a spinon Fermi surface, are not perturbatively accessible within our bosonic theory, and a complementary approach with fermionic partons is more suited in that scenario.
Also, detailed thermodynamic studies of the onset of field-induced order in the delafossites will shed light on the confinement transition.

  On the methodological front, it is worth pointing out different approaches from the literature. Spin-wave theory implemented for spins $1/2$ leads to a hard-core boson representation which has also been used to study magnetization plateau states and proximate phases \cite{misguich01,jolicoeur02,tay10}. Interestingly, the fractionalization schemes greatly differ: In our Schwinger-boson theory the spin fluctuations are composed of two bosonic partons. In the hard-core boson description, a spin flip corresponds to a single bosonic mode which, however, is then converted into a fermion via a Chern-Simons gauge field  \cite{misguich01,jolicoeur02} or fractionalizes into two fermionic partons~\cite{tay10}. Notably, a refined mean-field theory using hardcore bosons on the triangular lattice has found the same high-field sequence of transitions around the up-up-down phase~\cite{tay10} that we report.

\begin{acknowledgments}
  We thank P. M. C\^{o}nsoli and M. Protter for discussions and valuable inputs. This research was enabled in part by support provided by the Digital Research Alliance of Canada (\url{https://alliancecan.ca}). S.D.~was supported by the Faculty of Science at the University of Alberta. J.M.~was supported by NSERC Discovery Grants No. \#RGPIN-2020-06999 and No. \#RGPAS-2020-00064; the Canada Research Chair (CRC) Program; the Government of Alberta's Major Innovation Fund (MIF); the Tri-Agency New Frontiers in Research Fund (NFRF, Exploration Stream); and the Pacific Institute for the Mathematical Sciences (PIMS) Collaborative Research Group program. M.V.~acknowledges financial support from the DFG through SFB 
1143 (project-id 247310070) and the W\"urzburg-Dresden Cluster of Excellence on 
Complexity and Topology in Quantum Matter -- \textit{ct.qmat} (EXC 2147, project-id 390858490).
\end{acknowledgments}

\appendix

\section{Schwinger-boson mean-field theory on the triangular lattice}

In this appendix, we provide details regarding the large-$N$ (mean-field) Sp(N) Schwinger-boson formalism for investigating three-sublattice magnetic orders on the triangular lattice (Appendix~\ref{app:3sub}); we show how the classical TLHAF Hamiltonian is recovered in the $\kappa\rightarrow\infty$ limit (Appendix~\ref{app:classical}), and we provide explicit expressions for the trial condensate wave functions that encapsulate coplanar and non-coplanar orders (Appendix~\ref{app:ansatze}) and are used in our numerical solution of the mean-field equations (\ref{MF1}-\ref{MF3}).

\subsection{Three-sublattice structure}
\label{app:3sub}

To describe both QSL and magnetically ordered phases on the triangular lattice, we set up the formalism using a three-site unit cell, as this covers the semiclassical ordered states known for the TLHAF in a magnetic field~\cite{pifluxnote2}. We consider unit cells composed of non-overlapping upward-triangular plaquettes as depicted in Fig.~\ref{lattice}(a), with the sublattice labels $r=A, B, C$, the unit-cell (Bravais) translation vectors
\begin{align}\label{bravais}
\b{e}_1 &=\sqrt 3(0,1),\nn\\
\b{e}_2 &= \sqrt 3(-\sqrt 3/2,-1/2),\nn\\
\b{e}_3 &= \sqrt 3(\sqrt 3/2,-1/2),
\end{align}
and primitive translation vectors $\b{u}_1 = (1,0)$, $\b{u}_2 = (-1/2,\sqrt 3/2)$, and
$\b{u}_3 = (-1/2,-\sqrt 3/2)$. On going from the site basis ($i$) to the unit-cell basis ($\b{R},r$), where $\b{R}$ is a Bravais lattice vector (integer linear combination of the translation vectors (\ref{bravais})), the momentum-space spinon annihilation operator $b_{\b{k}ra\sigma}$ appearing in Eq.~(\ref{nambu}) is defined as
\begin{align}
    b_{\b{k}ra\sigma}=\sqrt{\frac{3}{\c{N}}}\sum_{\b{R}}e^{-i\b{k}\cdot\b{R}}b_{\b{R}ra\sigma}.
\end{align}

\begin{figure}
  \includegraphics[width=0.8\columnwidth]{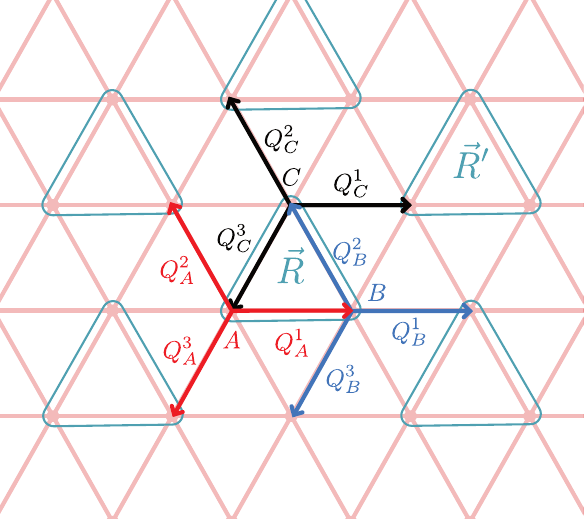}
  \caption{Three-sublattice structure of the fluctuating bond order-parameter $Q_r^\delta(\b{R})$ used in the computation of the
  Schwinger boson mean-field theory and its $1/N$ corrections.}
  \label{fig:Qfluc}
\end{figure}

For any three-sublattice bond configuration with a maximum of 9 independent bonds per unit cell, we use 
the parametrization
\beq
Q^\delta_r(\b{R})=Q_{r,r+\delta}, && \mu_r(\b{R}) = \mu_r,
\eeq
where $r\in\{A,B,C\}$, $\delta\in\{1,2,3\}$, and $r+\delta\in\{A,B,C\}$ denotes the sublattice index 
of the neighboring site along the primitive vector $\b{u}_\delta$ if $r$ is the sublattice index 
of the site $i=(\b{R},r)$ (Fig.~\ref{fig:Qfluc}). Here $Q_{r,r+\delta}$ and $\mu_r$ 
are the constant, uniform mean-field 
bond variables obtained at the given saddle point.

\begin{figure}
  \subfloat[]{
    \includegraphics[width=0.47\columnwidth]{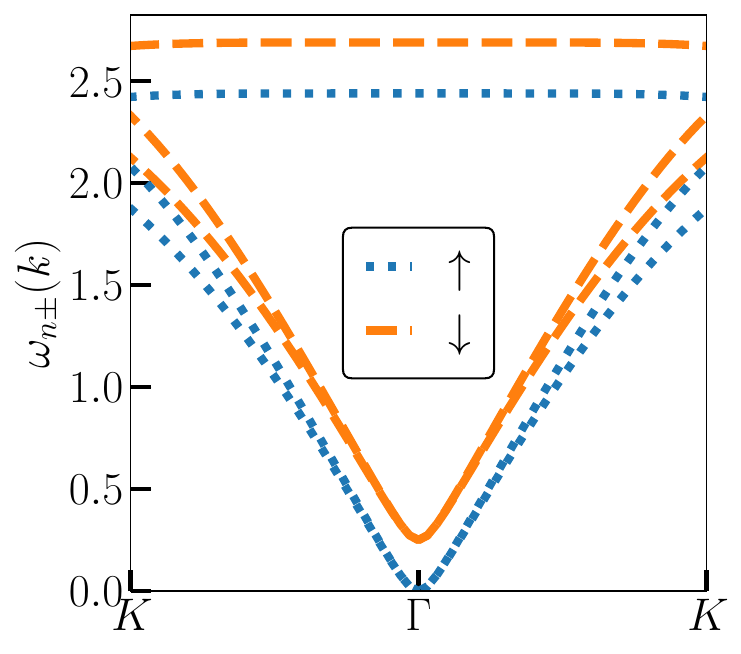}
  }
  \hfill
  \subfloat[]{
    \includegraphics[width=0.45\columnwidth]{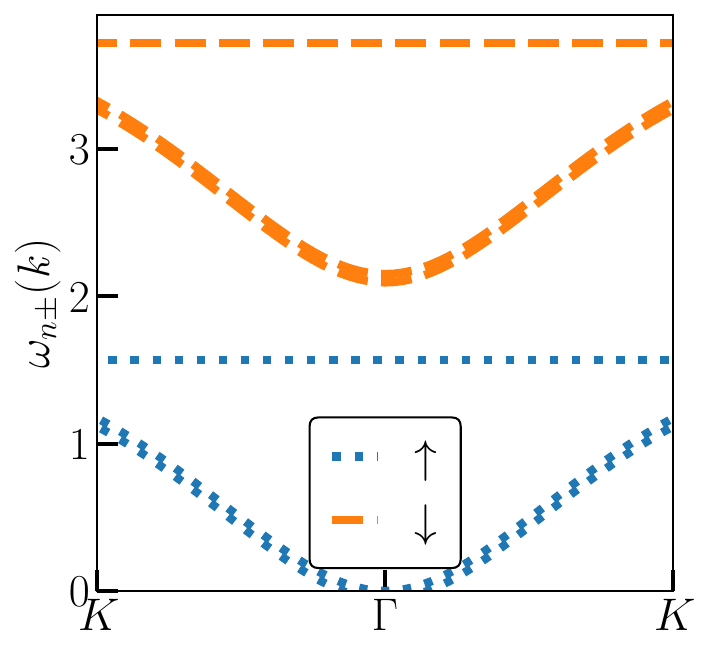}
  }
  \caption{Mean-field spinon spectrum in spinon BEC phases, plotted along the 
  K--$\Gamma$--K
  high-symmetry line in the reduced Brillouin zone, and computed for $\kappa=1$ in (a) the Y state ($h=0.50$) and (b) the UUD state ($h=4.30$).}
  \label{cnd_spec}
\end{figure}

For a generic and dynamic bond profile,
the quadratic Schwinger boson action is given by:
\beq
    \mathcal S_{b,b^*}
    =&\sum_{\b{R}ra}
    \int^\beta_0 d\tau\,
    b^*_{\b{R}ra\alpha}(\tau)\Big[
      \left(\partial_\tau+\mu_r(\b{R},\tau)\right)
      \delta_{\alpha\beta}\\
      &-
      \left(\frac{h\kappa}{4}+\frac{B_z}{2}\right)\sigma^z_{\alpha\beta}
      \Big]
    b_{\b{R}ra\beta}(\tau)
    \\
    &-\frac{1}{2}\sum_{\b{R}ra\delta}\int_0^\beta d\tau
    \Bigl(
      Q^\delta_r(\b{R},\tau)
      \\
      &\times\epsilon_{\alpha\beta}
      b^*_{\b{R}ra\alpha}(\tau)b^*_{\b{R}',r+\delta,a\beta}(\tau)
      +\mathrm{c.c.}\Bigr),\label{full_act}
\eeq
where $\b{R}'=\b{R}$ if the bond $(r,r+\delta)$ lies within unit cell $\b{R}$, and $\b{R}'\neq\b{R}$ if the bond connects $\b{R}$ to a neighboring cell $\b{R}'$ (Fig.~\ref{fig:Qfluc}). 
In the static limit, 
the lattice Fourier transformation of this action defines a pairing Hamiltonian,
\beq
&\hat{H}(
  \b{Q}_r(\b{p}),
  \b{Q}^*_r(-\b{p}),
  \mu_r(\b{p}),
  \b{k}+\b{p},\b{k})\\
  =&
  \begin{pmatrix}
    {\rm diag}(\mu_r(\b{p}))&& -J\gamma(\b{Q}_r(\b{p}),\k+\b{p},\k)/2\\
    -J\gamma(\b{Q}_r(\b{p}),\k+\b{p},\k)^\dagger/2&&{\rm diag}(\mu_r(\b{p}))
\end{pmatrix}
        \label{nonlocHam}
\eeq 
that is non-local
in momentum.
With the
constant mean-field parameters, the dynamical matrix $\hat{H}(\b{k})$ 
appearing in the effective Hamiltonian (\ref{hmf}) is obtained as 
the reduction of this pairing Hamiltonian
with momentum-independent Lagrange multipliers 
${\rm diag}(\mu_r)={\rm diag}(\mu_A,\mu_B,\mu_C)$. The pairing matrix $\gamma(\b{Q}_r,\b{k}',\b{k})$ 
is given by
\beq
  &\gamma(\b{Q}_r,\b{k}',\b{k})\\
  =&\left(\begin{array}{ccc}
    0 & \b{Q}_A
    \cdot\b{f}_A(\b{k}) & -\b{Q}_C\cdot\b{f}^*_C(\b{k}') \\
        -\b{Q}_A\cdot \b{f}_A^*(\b{k}') & 0 & \b{Q}_B\cdot\b{f}_B(\b{k}) \\
        \b{Q}_C\cdot\b{f}_C(\b{k}) & -\b{Q}_B\cdot\b{f}_B^*(\b{k}') & 0
        \end{array}\right),
        \label{nonlocPairing}
\eeq
with $\b{Q}_r=\b{Q}_r(\b{k}'-\b{k})$ collecting the
independent
  bond parameters incident on each unit-cell site of sublattice index $r$ 
  with the associated phase factors
\beq
  \b{f}_A(\b{k})&=( 1, e^{i k_2}, e^{-i k_3} ),\\
    \b{f}_B(\b{k})&=( 1, e^{-i k_1}, e^{i k_3} ),\\
      \b{f}_C(\b{k})&=( 1, e^{i k_1}, e^{-i k_2} ),
\eeq
where $k_j\equiv\b{k}\cdot\b{e}_j$, $j=1,2,3$. For the uniform mean-field solutions $Q_{rs}$, the non-local
expression for the pairing matrix is redundant, but it is a useful expression for fluctuation calculations. 
Upon diagonalizing $\hat{H}(\b{k})$ via a pseudo-unitary Bogoliubov transformation as described in Sec.~\ref{sec:BEC}, we obtain a mean-field spinon spectrum consisting of six spin-split bands $\omega_{n\pm}(\b{k})$, $n=1,2,3$. Magnetic order triggered by the onset of spinon BEC occurs when the lowest band touches zero energy (Fig.~\ref{cnd_spec}).

\subsection{The classical limit}
\label{app:classical}

As mentioned in Sec.~\ref{sec:classical}, the $\kappa\rightarrow\infty$ limit leads to additional simplifications. 
There $Q$, $\mu$, and $Z^\dag Z$ scale as $\kappa$, such that the energy density (\ref{emfcond}) is dominated by terms of order $\kappa^2$, while contributions from the uncondensed modes scale as $\kappa$. In the limit $\kappa\to\infty$, the condensate amplitude fulfills the normalization condition $Z^\dagger Z = 3\kappa$ arising from the Schwinger-boson occupation constraint. As shown earlier \cite{sachdev92}, the energetics in this limit is identical to that of classical spin vectors of 
length $S^{\rm cl}=\kappa/2$. Defining the classical vectors $\b{S}_r = z^\ast_{r\alpha} \b{\tau}_{\alpha\beta} z_{r\beta}/2$ and using the saddle-point values $Q_{rs}= \epsilon_{\alpha\beta} z_{r\alpha} z_{s\beta}$ of the bond variables, the condensate part of the 
mean-field energy density (\ref{emfcond}) can be brought into the form
\begin{align}
\frac{\varepsilon_0(\kappa\rightarrow\infty)}{N} &= J \sum_{r<s} {\b{S}}_r\cdot{\b{S}}_s - \frac{S^{\rm cl}}{6} \b{h} \cdot \sum_r {\b{S}}_r \nn\\
&\phantom{=}+  \frac{2}{3}\sum_r \mu_r (|\b{S}_r|-S^{\rm cl}),
\end{align}
representing a classical TLHAF model with three-sublattice ordering symmetry in the presence of a magnetic field, augmented by a Lagrange multiplier term to impose a length constraint on the spin vector.

\subsection{Mean-field ans\"atze}
\label{app:ansatze}

To efficiently solve the mean-field equations (\ref{MF1}-\ref{MF3}) at finite $\kappa$, we employ a numerical strategy that starts with the $\kappa\rightarrow\infty$ classical solutions and iterates the equations until convergence. Modulo global spin rotations about the field axis ($\hat{\b{z}}$) and a local $U(1)$ gauge degree of freedom, condensate wave functions describing classical coplanar and non-coplanar orders, respectively, can be written in terms of four Euler angles, 
$\b{\vartheta}=(\theta_A,\theta_B,\theta_C,\phi)$:
\begin{equation}
  \begin{aligned}
    Z(\b{\vartheta})=
    \sqrt{\kappa}
    \begin{pmatrix}
      \cos\left(\frac{\theta_A}{2}\right)\\
      e^{\frac{i\phi}{2}}\cos\left(\frac{\theta_B}{2}\right)\\
      e^{i\phi}\cos\left(\frac{\theta_C}{2}\right)\\
        -\sin\left(\frac{\theta_A}{2}\right)\\
      -e^{\frac{i\phi}{2}}\sin\left(\frac{\theta_B}{2}\right)\\
      -e^{i\phi}\sin\left(\frac{\theta_C}{2}\right)
    \end{pmatrix},
  \end{aligned}
\end{equation}
with the angles describing various ordering patterns determined by the external field magnitude ($\tilde h = h/J <9$) as,
\beq
\b{\vartheta}_{\rm cop.} &= 
\begin{cases}
\begin{pmatrix}
  \pi
  \\ \cos^{-1}\left(\frac{\tilde h+3}{6}\right)
  \\ -\cos^{-1}\left(\frac{\tilde h+3}{6}\right)
  \\0
\end{pmatrix} \ \text{ for } \tilde h \leq 3,\\
\begin{pmatrix}
  -\cos^{-1}\left(\frac{\tilde{h}^2-27}{6\tilde{h}}\right)
  \\ \cos^{-1}\left(\frac{\tilde{h}^2+27}{6\tilde{h}}\right)
  \\ \cos^{-1}\left(\frac{\tilde{h}^2+27}{6\tilde{h}}\right)
  \\0
\end{pmatrix} \ \text{ for } 3 <\tilde{h} \leq 9,\\
\end{cases}\\
\b{\vartheta}_{\rm noncop.} &= \begin{pmatrix}
  \cos^{-1}\left(\frac{\tilde h}{9}\right)
  \\ \cos^{-1}\left(\frac{\tilde h}{9}\right)
  \\ \cos^{-1}\left(\frac{\tilde h}{9}\right)
   \\ \frac{2\pi}{3}
\end{pmatrix}.
\label{Znoncop}
\eeq
Under the mapping (\ref{SpinOrder}), these correspond to degenerate ground-state solutions of the classical TLHAF with magnetic field $\tilde{h}<9$ (in units of $J$), which obey the energetic constraint $\b{S}_A+\b{S}_B+\b{S}_C=\tilde{h}S\hat{\b{z}}/3$~\cite{kawamura85}. For a given magnetic field $h$ in the Sp(N) Hamiltonian (\ref{largeN_ham}) with $\kappa$ just above $\kappa_c$ of the
$\mathbb{Z}_2$ QSL transition, it was shown in Ref.~\cite{yoshioka91} that to leading order,
the uniform moment of the condensate is proportional to $\tilde h \propto  h-h_c$, where $h_c$ is the critical field of the transition. Keeping that in mind, we use as initial guesses condensate ans\"{a}tze [Eq.~(\ref{Znoncop})] parameterized with general $\tilde h$ and not $h$. In practice, we sample over a range of $\tilde h$ to guide the numerical solver in order to achieve convergence.


\section{$1/N$ corrections}

In this appendix, we provide details regarding the computation of $1/N$ corrections in the Schwinger-boson formalism. We present the form of the three-point interaction vertex (Appendix~\ref{vertex}) and the Schwinger-boson Green's function (Appendix~\ref{app:bosonGF}), which are the basic blocks for diagrammatic calculations. We then compute the Schwinger-boson polarization bubble (Appendix~\ref{fluctSelf}), from which we obtain the RPA propagator $\overline{D}_A$ for bond-order-parameter fluctuations, which itself gives a $1/N$ correction to the ground-state energy density (Appendix~\ref{freeCorrection}).

\subsection{Three-point interaction vertex}
\label{vertex}

Beyond the mean-field level, we need to include spatial and temporal fluctuations in the bond order parameters 
$Q_{ij}$. Using the formalism developed in Appendix~\ref{app:3sub}, we expand around the static saddle-point bond configurations,
\beq
Q^\delta_r(\b{R})&=Q_{r,r+\delta}+A_{r\delta+}(\b{R}),\\
{Q^\delta_r(\b{R})}^*&=Q_{r,r+\delta}^*+A_{r\delta-}(\b{R}),
\eeq
and label the fluctuation modes with a single index $I=(r\delta s)$ where $s\in\{\pm\}$. In frequency-momentum space, the interaction vertex has the concise form,
\beq
\mathcal S_{\rm int} =\sqrt{\frac{3}{\N}} 
\sum_{K,P}A_I(P)\Phi_a^\dagger(K+P)\Gamma^I(K+P,K)\Phi_a(K)
\eeq
that is 
expressed graphically in Fig.~\ref{vacc_pol} with a 
vertex function $\Gamma^I$ given by: 
\beq
  \Gamma^I(K+P,K)
  &=\Gamma^I(\b{k}+\b{p},\b{k})\\
  &=
  \delta_{I,r\delta+}\frac{\partial\hat H(\b{k}+\b{p},\b{k})}{\partial Q_r^\delta(\b{p})}
  +\delta_{I,r\delta-}\frac{\partial \hat H(\b{k}+\b{p},\b{k})}{\partial {Q_r^\delta(-\b{p})}^*}
\eeq
with the non-local Hamiltonian introduced in Eq.~\eqref{nonlocHam}.


\subsection{Schwinger-boson Green's function}
\label{app:bosonGF}

The spinon Matsubara Green's function $G_\Phi(K)$ in Eq.~(\ref{bosonGF}) is a basic building block of our diagrammatic computations. Having diagonalized the dynamical matrix $\hat{H}(\b{k})$ as in Eq.~(\ref{bogoliubov}), we can compute the matrix inverse in Eq.~(\ref{bosonGF}) and obtain:
\begin{align}
  [G_\Phi(-ik_0,\b{k})]_{ij} = \sum_{n,\alpha\in\{\pm\}}
\frac{M_i^{n\alpha}(\b{k}) M_j^{n\alpha}(\b{k})^*}
  {-(ik_0+h\kappa/4)\sigma^z_{\alpha\alpha}+\omega_{n\alpha}(\b{k})},
\end{align}
where we have set the probe field $B_z$ to zero for simplicity and
used the spin-split notation for the Bogoliubov eigenmodes 
introduced in Eq.~\eqref{Upsilon}.

However, in the spinon BEC phases, we have to be careful with the Green's functions
due to the closing of the spinon gap at $\k=\b{0}$ (Fig.~\ref{cnd_spec}). In the various loop sums that follow, the 
condensation of the mode $c=(m_c+)$ is reflected via an
extensive weight of the Bose-Einstein function,
$n_B(\omega_c)
=\langle 
\xi_{m_ca\uparrow}(\b{0})^\dagger
\xi_{m_ca\uparrow}(\b{0})
\rangle
=1/(\exp(\beta(E_{m_c}(\b{0})-h\kappa/4))-1),$
for the condensate mode $m_c$ where the spinon gap closes.

The weight of the condensate mode is obtained by solving the mean-field energy equations in the
presence of the condensate,
\beq
&\frac{1}{\N}\frac{\partial E_{m_c}(\b{0})}{\partial \mu_r}n_B(\omega_c)+
\frac{1}{\N}\sum_{n}\sum_{\b{k}}\frac{\partial E_n(\b{k})}{\partial\mu_r}
=\frac{1}{3}(\kappa+1),\\
&n_B(\omega_c)=\frac{\N}{3(\partial E_{m_c}(\b{0})/\partial\mu_r)}|z_{r\alpha}|^2.
\label{condFract}
\eeq 
The first equation is obtained by writing out the diagonalized Hamiltonian [Eq.~\eqref{hmf}]
in its second-quantized form and taking the derivatives with respect
to the Lagrange multipliers. In the last line, a comparison of the equation
with Eq.~\eqref{MF2} defines the weight of the Boson condensate at
zero temperature. 

In the Matsubara frequency sum involving the spinon Green's function, we introduce this normalization
factor for the condensate-mode Bose-Einstein function. Following Ref.~\cite{zhang22}, 
we consider a non-fragmented, simple BEC of only one of the possibly degenerate low-lying 
modes.

\subsection{Polarization bubble}
\label{fluctSelf}

Let us first consider the full dynamical content of the fluctuation theory of the Schwinger boson theory. 
To compute the $1$-loop polarization function,
we simplify its expression given in the main text in terms of the modified vertices,
\beq
\gamma^I(K+P,K)=M(K+P)^\dagger \Gamma^I(K+P,K) M(K),
\eeq
where $M(K)\equiv M(\boldsymbol k)$ are the Bogoliubov transformation matrices. With the modified vertices,
the loop-sum becomes
simpler,
\begin{widetext}
  \begin{align}
      \Pi_{\rm 1-loop}^{IJ}(P)&=\frac{3}{\beta\mathcal N}\sum_{K}
      [\gamma^I(K+P,K)]_{m\alpha,n\beta}[g_\Phi(K)]_{n\beta}[\gamma^J(K,K+P)]_{n\beta,m\alpha} 
      [g_\Phi(K+P)]_{m\alpha}\nonumber\\
       &=\frac{3}{\beta\mathcal N}\sum_{K}
       \pi^{IJ}_{m\alpha,n\beta}(K+P,K)[g_\Phi(K+P)]_{m\alpha}[g_\Phi(K)]_{n\beta},\nonumber\\
  \end{align}
\end{widetext}
where $\pi^{IJ}_{m\alpha,n\beta}(K+P,K)=[\gamma^I(K+P,K)]_{m\alpha,n\beta}
[\gamma^J(K,K+P)]_{n\beta,m\alpha}$. The expression involves
a Matsubara frequency summation and summation over momentum modes. 

The function $\pi^{IJ}$ only
involves momentum. 
The frequency summation, on the other hand,
can be computed exactly with the bosonic Matsubara frequency summation formula
(c.f. pp. 247 in Ref.~\cite{coleman15}) by replacing the sum with a contour integral around the poles
of the summand,
$
\frac{1}{\beta}\sum_{ik_0}f(ik_0) = -\int_{\mathcal C}\frac{dz}{2\pi i}f(z)n_B(z),
$ where $n_B(z)=1/(\exp(\beta z)-1)$ is the Bose-Einstein function. The 
Matsubara summation within our polarization function evaluates to
\begin{equation}
  \begin{aligned}
    &\frac{1}{\beta}\sum_{ik_0}[g_\Phi(K+P)]_{m\alpha}[g_\Phi(K)]_{n\beta}\\
    =&-
    \sigma^z_{\alpha\alpha}
    \sigma^z_{\beta\beta}
    \frac{
      n_B(\sigma^z_{\alpha\alpha}\omega_{m\alpha}(\b{k}+\b{p}))
      -n_B(\sigma^z_{\beta\beta}\omega_{n\beta}(\b{k}))
    }{-ip_0+
    \sigma^z_{\alpha\alpha}\omega_{m\alpha}(\b{k}+\b{p})
    -\sigma^z_{\beta\beta}\omega_{n\beta}(\b{k})
    }.
  \end{aligned}
\end{equation}
At zero temperature,
the Bose function
contributes to a macroscopic
weight of the density of states at the condensate mode momentum. Introducing the polarization
wave-function,
\beq
&L^{IJ}_{m\alpha,n\beta}(\k+\b{p},\k)
=\pi^{IJ}_{m\alpha,n\beta}(\b{k}+\b{p},\b{k})
 \sigma^z_{\alpha\alpha}
    \sigma^z_{\beta\beta}\\
    &\times
      \left(n_B(\sigma^z_{\alpha\alpha}\omega_{m\alpha}(\b{k}+\b{p}))
      -n_B(\sigma^z_{\beta\beta}\omega_{n\beta}(\b{k}))\right),
\eeq
and the two-spinon dispersion
$\mathcal E_{m\alpha,n\beta}(\k+\b{p},\k)=\sigma^z_{\alpha\alpha}\omega_{m\alpha}(\b{k}+\b{p})
    -\sigma^z_{\beta\beta}\omega_{n\beta}(\b{k})
$,
we evaluate the polarization function as a momentum summation over the first Brillouin zone,
  \begin{equation}
    \begin{aligned}
      \Pi^{IJ}(-ip_0,\b{p})
      =&-\frac{3}{\mathcal N}\sum_{mn\alpha\beta
      \boldsymbol k}
        \frac{L^{IJ}_{m\alpha,n\beta}(\boldsymbol k+\b{p},\boldsymbol k)}{-ip_0+
       \mathcal E_{m\alpha,n\beta}(\boldsymbol k+\b{p},\boldsymbol k)}.
       \label{polFunc}
    \end{aligned}
  \end{equation}

\subsection{Correction to the energy density}
\label{freeCorrection}

Loop corrections to free energy may formally diverge. To consider the finite part of the free energy
correction (\ref{ener-1/N}), we perform the following summation, 
\beq
\varepsilon_{0,\rm reg}^{(1)}
=&\frac{1}{2\beta\N}\sum_{ip_0,\b{p}}
\Big(
\tr\ln\left[D_A^{-1}-\Pi(-ip_0,\b{p})\right]\\
&-
\tr\ln\left[D_A^{-1}\right]
\Big),
\eeq
where we have subtracted a constant factor from the expression to regularize it. 
Due to the regularization, this is a finite expression that can be computed by using the 
contour-deformation technique for Matsubara frequency summation,
\beq
\varepsilon_{0,\rm reg}^{(1)}
=-\frac{1}{2\mathcal N}\sum_{\b{p}}\int_{\mathcal C}\frac{dz}{2\pi i}
\tr\ln\left[I-D_A\cdot\Pi(z,\b{p})\right]n_B(z),
\eeq
with the resulting complex integral evaluated with the residue theorem. 
The pole structure of the logarithm is
determined by possible branch-cut singularities for positive frequencies, $z\in[0,\infty)$, and the isolated regular poles
of the polarization function. We treat them separately.

\subsubsection{Contribution from isolated poles of the log}

Let us consider an integral path around an isolated singularity of the polarization function,
$
z=\mathcal E_{m\alpha,n\beta}(\boldsymbol k+\b{p},\boldsymbol k)+\rho e^{i\theta},
$
where $\rho\rightarrow 0$ 
is an infinitesimal radius around the singularity. Around that pole, the above contour integral becomes, 
\beq
&\lim_{\rho\rightarrow 0}
-\frac{1}{2\mathcal N}\int \frac{i\rho e^{i\theta} d\theta}{2\pi i}
\Bigg(
\tr\ln\big[I-\frac{3}{\mathcal N}\frac{D_A\cdot\hat L_{m\alpha,n\beta}(\boldsymbol k+\b{p},\boldsymbol k)}{\rho 
  e^{i\theta}}\\
  &+\dots\big]
  \Bigg)n_B(\mathcal E_{m\alpha,n\beta}(\k+\b{p},\k)),
\eeq
where $[\hat L]^{IJ}=L^{IJ}$. Now, by expanding the log, it is easy to see that the only non-zero contribution in this
expression comes from the first-order term, as all the higher-order $O(1/\rho^2)$ terms cancel with
their associated angular
integrals vanishing identically. 
Now, by summing over contributions from all these poles we obtain the first contribution to the
regularized free-energy,
\beq
&\varepsilon_{0,\rm reg}^{(1)}(\text{poles})\\
=&\frac{3}{2\N^2}\sum_{mn\alpha\beta,\boldsymbol k,\b{p}}
\tr[D_A\cdot\hat L_{m\alpha,n\beta}(\boldsymbol k+\b{p},\boldsymbol k)]\\
&\times
    n_B(\mathcal E_{m\alpha,n\beta}(\boldsymbol k+\b{p},\boldsymbol k)),
\eeq
where the final sum over the first Brillouin zone lacks an analytical expression and has to be performed
numerically. 

\subsubsection{Contribution from branch-cut singularities of the log}

A branch-cut singularity is encountered when, at a special pole $z_{\rm br}=\mathcal E_{m\alpha,n\beta}(\b{k}+\b{p},\b{k})$, 
the argument of the logarithm vanishes. This implies that the eigenvalue equation,
\beq
\left[I-D_A\cdot\Pi(z_{\text{br}},\b{p})\right]|v_\lambda\rangle=\lambda
|v_\lambda\rangle,
\eeq
has at least one zero eigenvalue for branch cuts ending at poles. 
Let us 
introduce a short-hand notation, $\mathcal E_\ell=\mathcal E_{m\alpha,n\beta}(\k+\b{p},\k)$
for the two-spinon poles. The 
additional contribution from the branch cut associated with the special pole evaluates
to
\beq
&\varepsilon^{(1)}_{0,\rm reg}(\text{branch})\\
=
&-\frac{1}{2\mathcal N}\sum_{\b{p}}
\int_{\mathcal E_{\ell}}^\infty
\frac{dw}{2\pi i}
\tr\Big[\ln\left[I-D_A\cdot\Pi(w-i\eta,\b{p})\right]\\
&-\ln\left[I-D_A\cdot\Pi(w+i\eta,\b{p})\right]
n_B(w),
\eeq
with the selection
of a counter-clockwise contour shifted $\pm\eta$ from the
positive-real frequency axis. By taking the limit  $\eta\rightarrow 0$, it is easy to see that
the surviving contribution in this equation comes from the locus in the (first) reduced
Brillouin zone with $\mathcal E_\ell=0$.

Following some more algebra, the branch-cut contributions are enumerated to be
\beq
&\varepsilon^{(1)}_{0,\rm reg}(\text{branch})\\
=&
-\frac{1}{2\mathcal N}\sum_{\b{p}}\sum_{\ell}
\delta_{\mathcal E_{\ell},0}
\int_0^\infty
\frac{dw}{2\pi i}n_B(w)\\
&\times\sum_{\lambda^\ell}
\left(
\ln\left[1+\frac{3}{\N}\frac{\lambda^\ell}{-\omega+i\eta}\right]
-
\ln\left[1+\frac{3}{\N}\frac{\lambda^\ell}{-\omega-i\eta}\right]
\right)
\\
=&\frac{1}{2\beta\N}\sum_{\b{p},\ell}\delta_{\mathcal E_\ell,0}\sum_{\lambda^\ell}
\ln\left(1-\exp(-3\beta\lambda^\ell/\N)\right),
\eeq
where $\lambda_\ell$ are the eigenvalues of the spectral decomposition,
\beq
D_A\cdot\hat L_{\ell} =\lambda^{\ell}|v_{\lambda^{\ell}}\rangle\langle
v_{\lambda^{\ell}}|,
\eeq
with $[\hat L_{\ell}]^{IJ} = L^{IJ}_{m\alpha,n\beta}(\k+\b{p},\k)$ being the polarization wave-function defined
above and re-expressed in the short-hand notation. This contribution vanishes at zero temperature, 
$\beta\rightarrow\infty$.


\section{Spin structure factor}
\label{SpinStruct}

An experimentally relevant observable is the dynamic spin
structure factor,  Eq.~(\ref{Sqw}), where $S_i^\alpha(t)=e^{iHt}S_i^\alpha e^{-iHt}$ are the
real-time Heisenberg-picture spin operators, $i$ is a site index, and $\alpha,\beta$ are spin indices. This is a
real-time four-spinon correlator that can be computed from our Euclidean theory by using analytic
continuation~\cite{sachdev11},
\beq
S^{\alpha\beta}(\omega,\b{p})=(\Theta(\omega)/\pi)\text{Im}
\left[\chi^{\alpha\beta}(\omega+i\eta,\b{p})\right]_{\eta\rightarrow 0},
\eeq
where $\chi^{\alpha\beta}(i\omega,\b{p})$
is the spin susceptibility and we replaced $i\omega\mapsto \omega+i\eta$ to obtain
the real-time response function.
The spin-susceptibility is obtained by considering our original action with the Zeeman-coupling
source-term $\boldsymbol B$ [Eq.~\eqref{SBact}], 
and then taking the derivative concerning this field \cite{zhang22}
,
\beq
\chi^{\alpha\beta}(P) = \frac{\partial^2\ln Z(B)}{\partial B_\alpha(P)
\partial B_\beta(-P)}.
\eeq
The result of the functional derivative can be organized in terms of a diagrammatic $1/N$
expansion but with some subtleties. 

\begin{figure}[bt]
  \includegraphics[width=0.95\columnwidth]{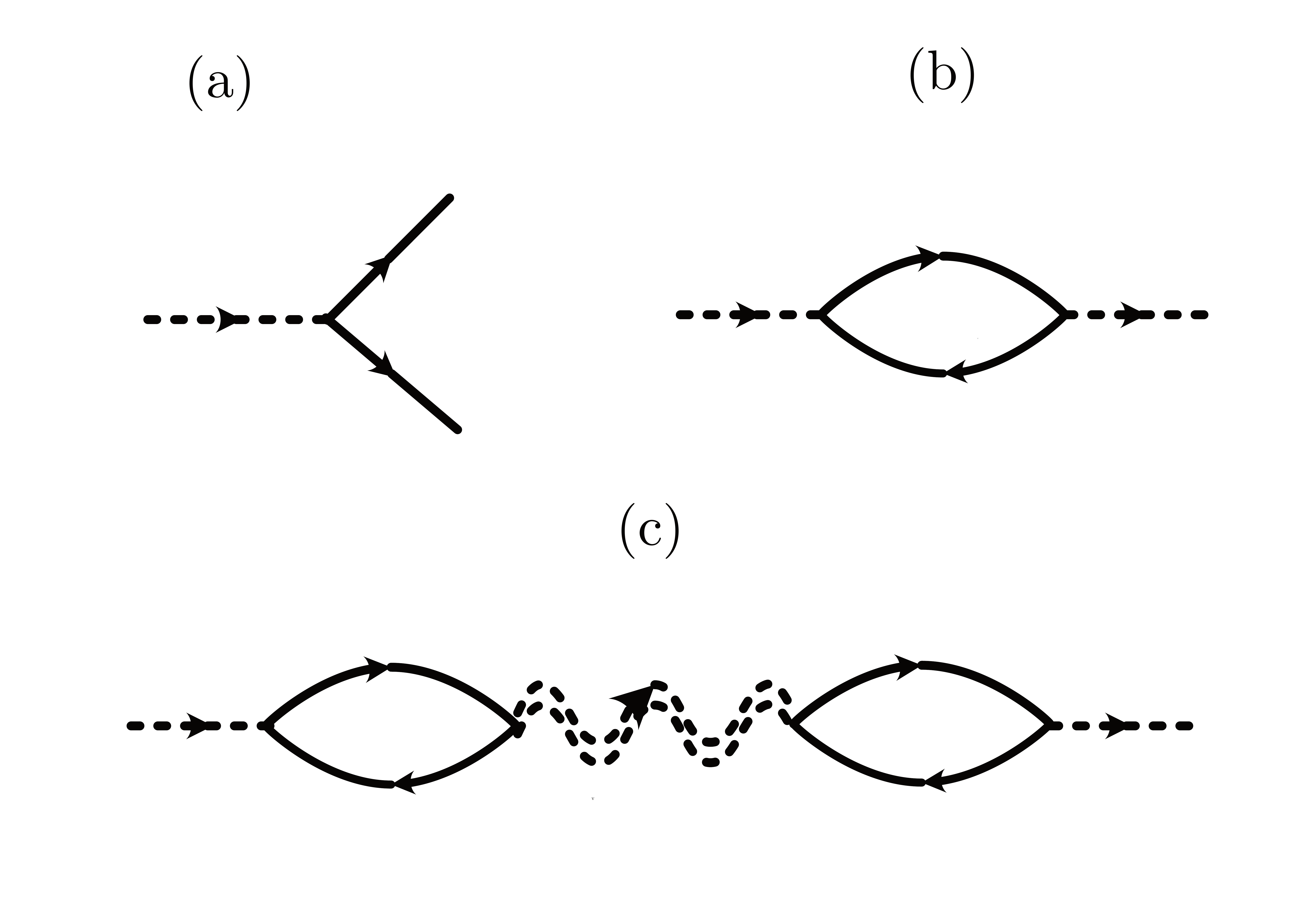}
  \caption{The Feynman diagrams for the approximated susceptibility
  computation; (a)
  the Zeeman vertex, (b) the leading-order spin-susceptibility loop,
  and (c) a subleading correction to the spin-susceptibility loop with dressed fluctuation propagator.}
  \label{sfcfeyn}
\end{figure}

First of all, to accommodate all the spin vertices, we need to double our Nambu basis,
\beq
\Psi_a(K)&=
\begin{pmatrix}
  b_{Aa\uparrow}(K)\\
  b_{Ba\uparrow}(K)\\
  b_{Ca\uparrow}(K)\\
  b_{Aa\downarrow}(K)\\
  b_{Ba\downarrow}(K)\\
  b_{Ca\downarrow}(K)\\
  b^*_{Aa\uparrow}(-K)\\
  b^*_{Ba\uparrow}(-K)\\
  b^*_{Ca\uparrow}(-K)\\
  b^*_{Aa\downarrow}(-K)\\
  b^*_{Ba\downarrow}(-K)\\
  b^*_{Ca\downarrow}(-K)\\
\end{pmatrix},
\eeq
with the corresponding Green's function $2 G_\Psi(K)$.
Furthermore, we need to connect the lattice momentum of the probe Zeeman field
to the Brillouin-zone momentum of our enlarged unit cell;
\beq
&\delta H[\boldsymbol B] =\sum_{a,i} \boldsymbol B_i\cdot b^\dagger_{a,i,\alpha}
[\boldsymbol\sigma/2]_{\alpha\beta}b_{a,i,\beta}\\
=&\frac{1}{\sqrt{\N}}\sum_{\boldsymbol x}
\sum_{\b{p}} \boldsymbol B (\b{p})\cdot e^{i\b{p}\cdot(\boldsymbol x+\boldsymbol\delta_r)}
b^\dagger_{ar\alpha}(\boldsymbol x)[\boldsymbol\sigma/2]_{\alpha\beta}b_{ar\beta},
\eeq
where $\boldsymbol\delta_r$ represent the displacement vectors within a unit cell, e.g.,
$\boldsymbol\delta_A = (0,0)$, $\boldsymbol\delta_B=(1,0)$, and $\boldsymbol\delta_C=(1/2,\sqrt 3/2)$,
where we have placed the coordinate of a unit cell at its A-sublattice site. In the enlarged Nambu basis, the spin vertices are now given by 
[Fig.~\ref{sfcfeyn}~(a)] $\delta\mathcal S[B]=-(1/2\sqrt{\beta\mathcal V})
\sum_{Q,K} \Psi^*_a(K+P)[\boldsymbol B(P)\cdot\boldsymbol J(P)]\Psi_a(K)$, where the vertices
$\boldsymbol J(P)$ can be read off from the equation above.

The leading-order large-$N$ contribution to the dynamic susceptibility is given by Fig.~\ref{sfcfeyn}~(b),
\beq
\chi^{\alpha\beta}(P) =
\frac{1}{\beta \mathcal V}\sum_{K}\Tr\left[J^\alpha(P) G_\Psi(K)
J^\beta(-P)G_\Psi(K+P)\right].
\eeq
Due to the presence of a gap closing the boson spectrum, the spinon Green's functions are singular at the momentum $\boldsymbol k = 0$. This singularity
is related to the Bose-Einstein condensate through the condensate Green's function. 

\begin{figure}
  \subfloat[]{
    \includegraphics[width=0.48\columnwidth]{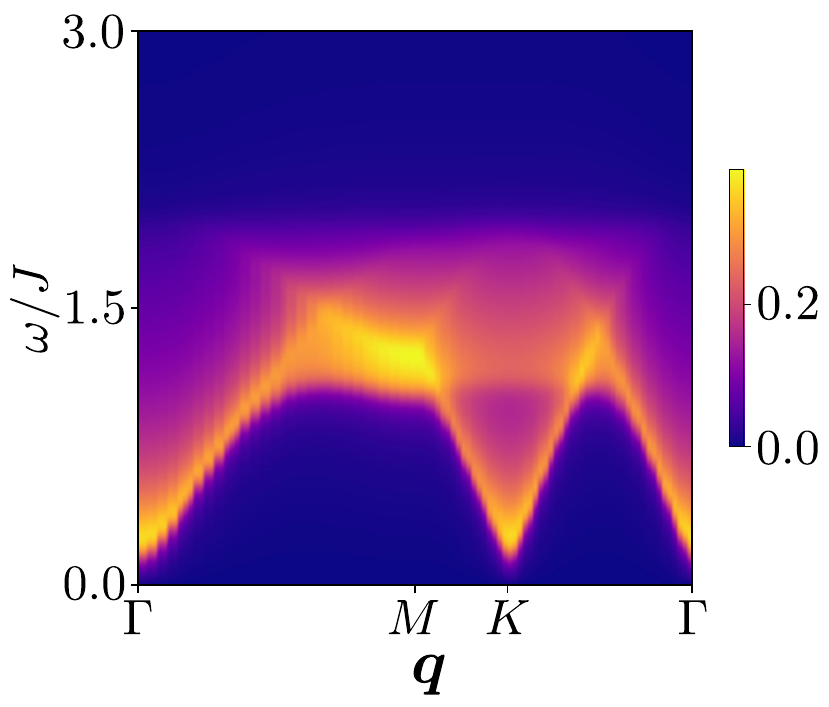}
  }
  \subfloat[]{
    \includegraphics[width=0.48\columnwidth]{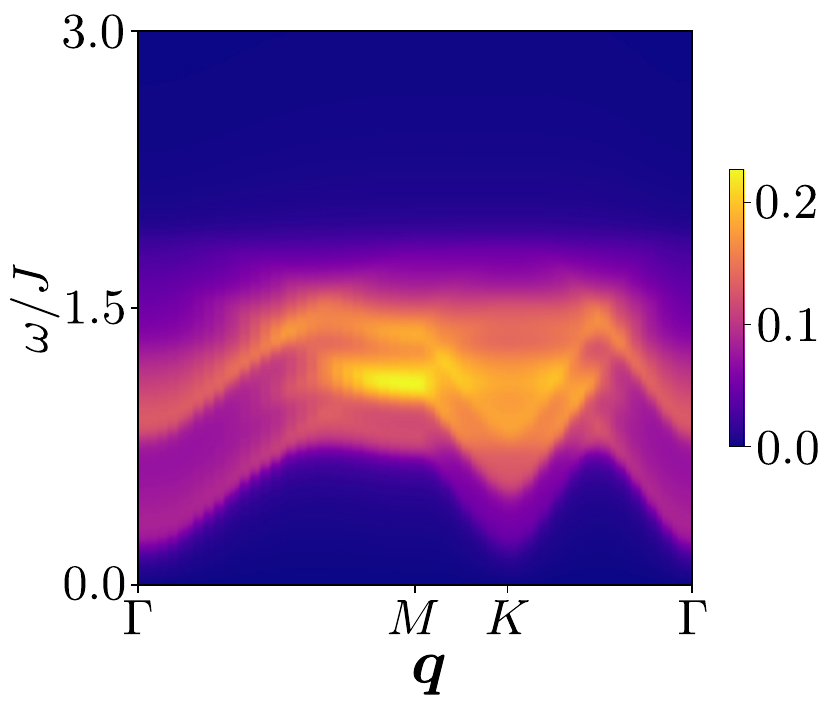}
  }
  \caption{The isotropic dynamic structure factor $S(\omega,\b{q})$ plotted along
  the high-symmetry points of the triangular lattice for (a) $\kappa=0.3, h = 0.1 J$
  and
  (b) $\kappa=0.2, h = 3 J$ in 
  the spin liquid phases of the Schwinger boson theory.
  }
  \label{structureExtra}
\end{figure}

In the disordered phase, the spin structure factor is
dominated by a two-spinon continuum characteristic of a gapped spin liquid
[Fig.~\ref{structureExtra}~(a)]. The situation becomes more complicated in the
ordered phase.
It is straightforward to see that the presence of the condensate leads to a 
single-spinon-pole spectral response in the leading-order spin susceptibility. The
leading-order mean-field theory, therefore, leads to spurious spin dynamics.
References~\onlinecite{ghioldi18,zhang22} showed that this artifact of the leading
large-$N$
expansion can be controllably eliminated by considering diagrams that appear to be
higher order in $1/N$ but contributes to lower orders due to the presence of the
condensate. The higher-order sister diagram of Fig.~\ref{sfcfeyn}(b) is
Fig.~\ref{sfcfeyn}(c) which includes the fully-dressed RPA propagator for the
fluctuations and yields single spinon poles with precisely the opposite sign
and cancels the spurious pole from the leading-order diagram \cite{zhang22}.
Without going into the specifics of these calculations, we show the resulting
isotropic spin structure factor, $S(\omega,\boldsymbol k)
=\sum_\alpha S^{\alpha\alpha}(\omega,\boldsymbol k)$, for $\kappa=1$ in Fig.~\ref{structure}.
In Fig.\ref{structureExtra}~(b), the spinon gap is not yet closed for $\kappa=0.2$ and
$|\boldsymbol h|/J = 3$ but it is small.
In Fig.~\ref{structureExtra}, we can see the two-spinon continuum that
is a hallmark of the spin-liquid phase, but, the
spectra are spin-split due to the presence of the
Zeeman field.

\bibliography{triangsb_long}

\begin{thebibliography}{62}
\bibitem [{pif()}]{pifluxnote2}%
  \BibitemOpen
  \href@noop {} {}\bibinfo {note} {The $\pi$-flux QSL solution would require to
  consider a two-site unit cell. However, it is known to have a higher energy
  than the zero-flux state~\cite{wang06} and is therefore not considered
  here.}\BibitemShut {Stop}%
\end{thebibliography}

\end{document}